\RequirePackage{fix-cm}
\documentclass[smallextended]{svjour3}       
\smartqed  
\usepackage{graphicx}
\usepackage{cite}
\usepackage{float,multirow,amsmath,amssymb,cuted,flushend,multicol}
\usepackage[colorlinks,citecolor=blue,urlcolor=blue,linkcolor=blue]{hyperref}
\usepackage{authblk}
\begin{document}

\title{Quasi Normal Modes of Ayon-Beato Garcia Regular Black Holes For Scalar Field}



\author{Masum Murshid         \and
        Farook Rahaman \and
        Mehedi Kalam}


\institute{F. Author \at
              Department of
Physics, Aliah University, II-A/27, Action Area II, Newtown,
Kolkata-700156, India.\\
              \email{masum.murshid@wbscte.ac.in}           
           \and
           S. Author \at
              Department of Mathematics, Jadavpur University, Kolkata 700032, West Bengal, India.\\
              \email{rahaman@associates.iucaa.in}
              \and
              T. Author \at
              Department of
Physics, Aliah University, II-A/27, Action Area II, Newtown,
Kolkata-700156, India.\\
              \email{kalam@associates.iucaa.in} 
}

\date{Received: date / Accepted: date}

\maketitle

\begin{abstract}
In this paper, we compute quasi-normal modes of ABG black holes (which has a non-linear electrodynamical source) using the WKB methods and AIM. A comparison between the spectrum of QNMs
calculated by both methods is made. We analyse how the spectrum
of QNMs depends on the black hole parameters, multipole number
and overtone number and establish that the ABG black hole is stable against the scalar field.
\end{abstract}

\section{Introduction} \label{Introduction}
Black Holes(BH) are the most radical prediction of Einstein's
General relativity. The study of BH had started in the early
20th century. BH is a very simple object when it's kept in isolation,
and these isolated BHs are characterized by only a few parameters
such as their mass, charge and angular momentum. However, a BH
cannot be isolated because it interacts with the surrounding
astrophysical objects such as accretion disk, stars, planets etc
or fields such as scalar field, electromagnetic field, Weyl
field. Thus a BH is always in the perturbed state. The perturbed
BH responds to perturbations by emitting gravitational waves.
The dynamical evolution of these gravitational waves can be
conditionally divided into three-stage. Among three divisions of
the dynamical evolution of gravitational waves, the damping
oscillations of black holes have great interest, known as Quasi-
Normal Modes(QNMs) of Black Hole. These QNMs are the dependent
of the parameters of BH like mass, charge and angular momentum,
of the types of the perturbation like a scalar, electromagnetic
etc which are the cause of the excitation of BH. The frequencies
of these QNMs are complex, where the real part gives the
frequency of the perturbation and the imaginary part gives its
damping rate. Since these QNMs depend on the BH parameters, it
provides us with lots of information about BH. QNMs are
extensively used in conjecture Ads/CFT correspondence
\cite{Maldacena1999}. The return to thermal equilibrium of the
perturbed state in CFT is related to the imaginary part of the
QNMs \cite{PhysRevD.62.024027}. QNMs are also used to interpret
loop quantum gravity \cite{BEKENSTEIN19957,
PhysRevLett.81.4293,PhysRevLett.90.081301} and to test different
alternative gravity theories.   
     \par  Regge and Wheeler \cite{Regge:1957td} were the first persons who discovered that the whole study of perturbation theory of BH could be
reduced to a Schr{\"o}dinger-like equation and their work was
extended by Zerilli \cite{PhysRevD.2.2141,PhysRevLett.24.737} .
In 1970, Vishveshwara \cite{PhysRevD.1.2870} studied the
stability of Schwarzschild metric and showed QNMs behaviour of BH
by scattering of gravitational waves \cite{Vishveshwara:1970zz}.
Chandrasekhar and Detweiler proved that the Regge-Wheeler and
Zerilli potential are isospectral and also calculated QNMs for
Schwarzschild BH, using shooting method
\cite{Chandrasekhar:1975zza}. Hans-Joachim Blome and Bahram
Mashhoon \cite{BLOME1984231} analytically calculated QNMs of
Schwarzschild BH states of the inverted effective potential of
BH, an approximation of Eckart potential. In the same year,
Ferrari and Mashhoon \cite{PhysRevD.30.295} used the same
technique to compute QNMs but this time they used
P{\"o}shi-Teller potential to approximate inverted BH effective
potential. Leaver published series of paper
\cite{doi:10.1063/1.527130,PhysRevD.34.384,Leaver:1985ax} to
calculate QNMs numerically using Continuous Fraction Method (CFM)
. Later on some new methods like Horowitz Hubeny method (HH)
\cite{Horowitz-Hubeny:2000}, asymptotic iteration method (AIM)
\cite{Cho_2010} were proposed to calculate QNMs numerically.
Some review papers \cite{Nollert:1999ji,
Kokkotas1999,Berti_2009, Konoplya:2011qq} give a brief
description of QNMs and it's applications. 
The modified classical general relativity or quantum gravity may give rise to a class of BH with no pathological singular behaviour. These singular BHs are called regular BH \cite{Bardeen,Mars_1996,Barrab,Borde,AyonBeato:1998ub,Ayon-Beato1999,Ayon-Beato:2000mjt,PhysRevLett.96.031103}. These regular BHs can be constructed by coupling gravity to matter having non-linear electrodynamic behaviour. Flachi and Lemos \cite{PhysRevD.87.024034}, Toshmatove et al. \cite{PhysRevD.91.083008,Toshmatov:2018ell}, Wu  \cite{refId0} and several other authors like \cite{Li:2016oif,Lopez:2018aec,Panotopoulos:2019qjk,PhysRevD.103.124050} found QNMs of regular BHs. Hendi et al. \cite{PhysRevD.103.064016} studied the thermodynamic properties of the regular rotating  BH using QNMs. Cai and Miac \cite{Cai:2020kue} studied the regular BH in scalar-tensor-vector gravity. Bronnikov et al. \cite{Bronnikov:2012ch} showed that the regular BHs and wormholes are unstable against the phantom scalar field except for one special class of black universe. Churilova and Stuchlik \cite{Churilova:2019cyt} studied the ringing of the one-parameter family of static regular BH/Wormhole geometries and showed that the BH/Wormhole transition is characterised by echoes. Bronniko and Konoplya \cite{PhysRevD.101.064004} studied the Echoes in Brane World that occurred during the BH/wormhole transition.

\par In this paper, we discuss regular BH in section-\ref{Formation} and then we discuss dynamical behaviour of the scalar field in section-\ref{Dynamical}. In section-\ref{Methods}, we talk about two methods of calculation of QNMs( which are being applied in this paper ). After that, we present some calculation which is required to perform AIM in section-\ref{cal}. Finally, in section-\ref{conclusion}, we discuss some of the results of this paper.
\par In this paper, all quantities are expressed in the natural unit. All the numerical calculations are done by substituting the mass of BH $M=1$ and the mass of the scalar field $\mu=0$.
\begin{figure*}[!ht]
    \centering
        \begin{minipage}[b]{0.45\textwidth}
            \includegraphics[width=1\textwidth]{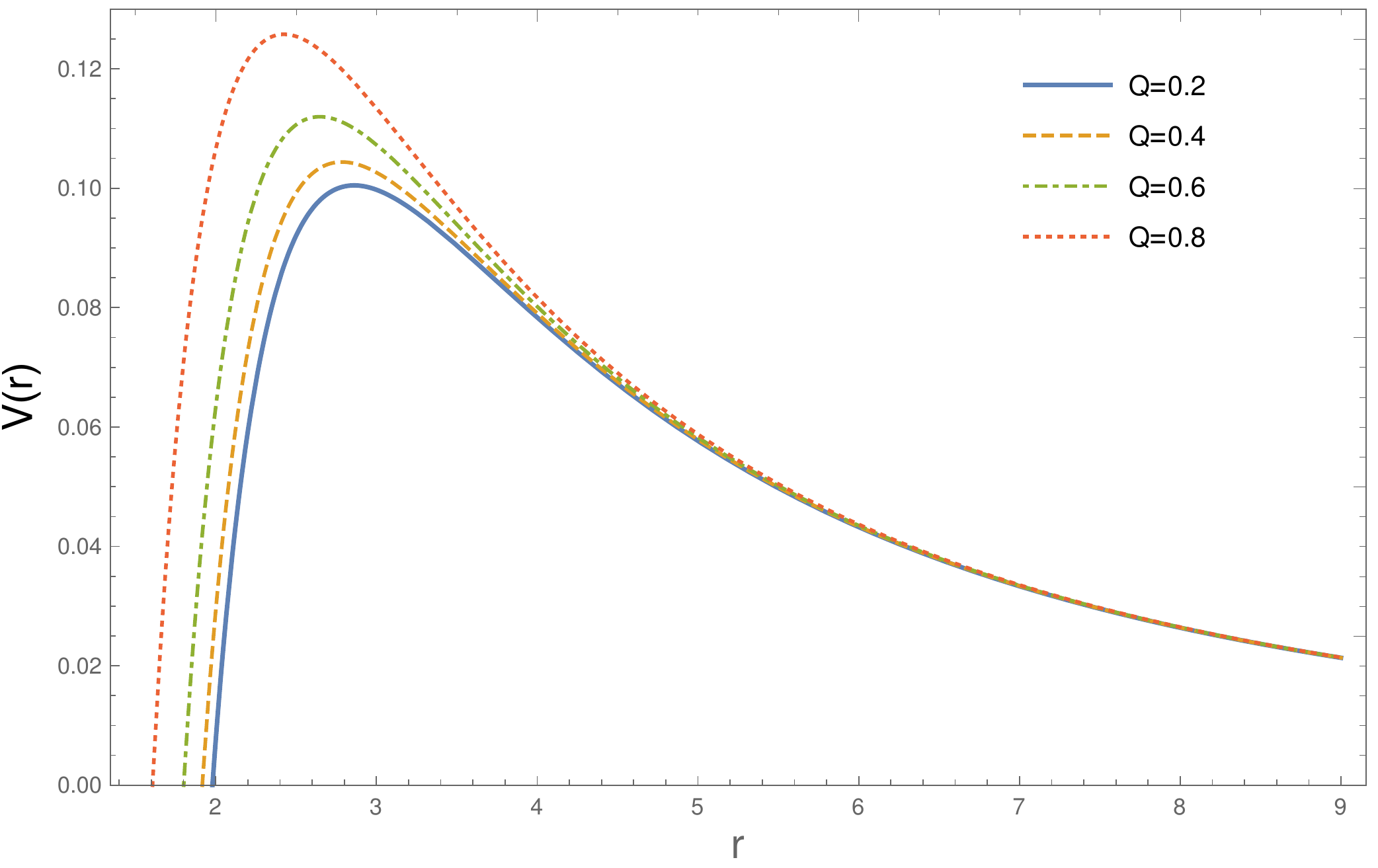}
        \end{minipage}\qquad
        \begin{minipage}[b]{0.45\textwidth}
            \includegraphics[width=1\textwidth]{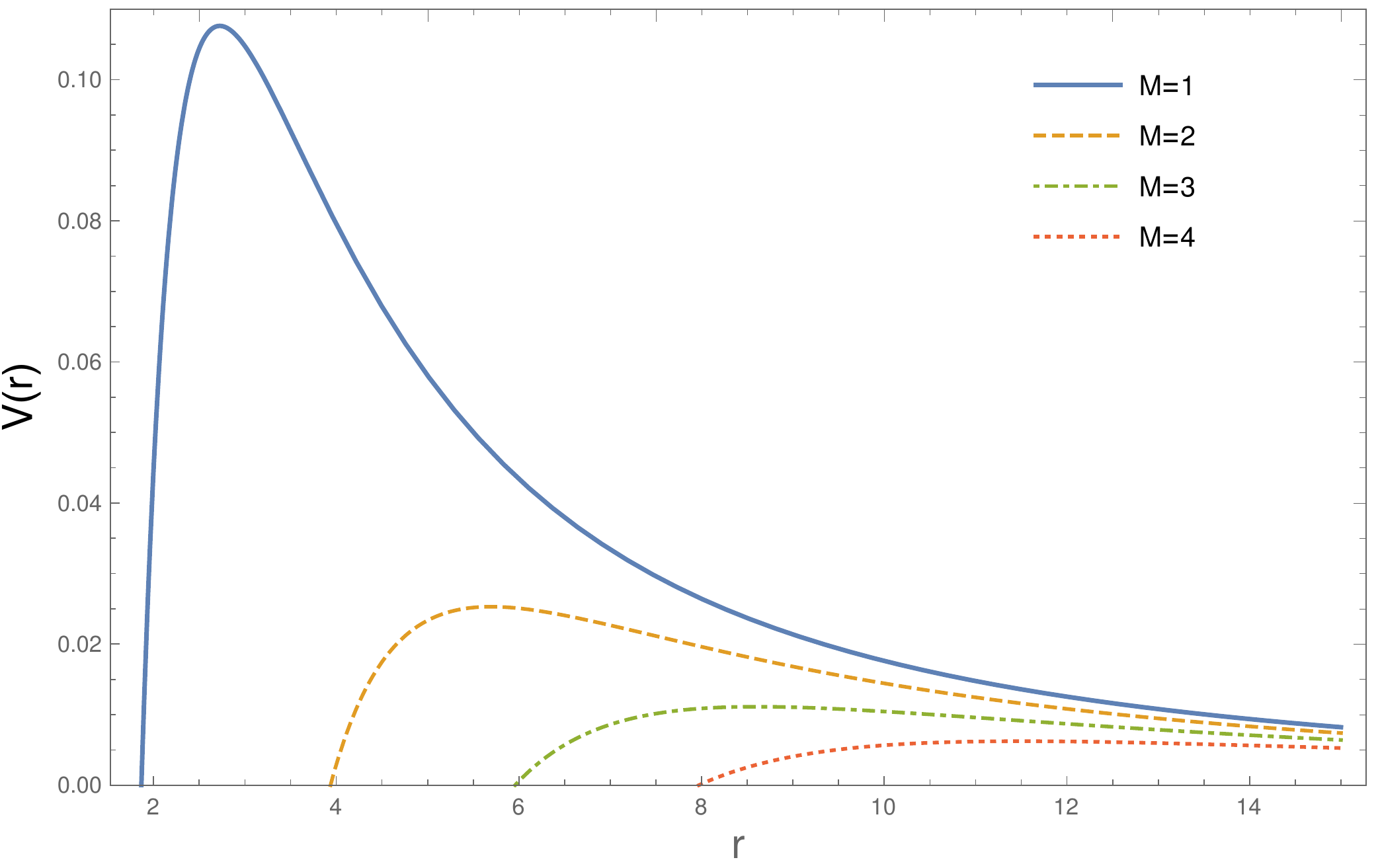}
        \end{minipage}\qquad
        \begin{minipage}[b]{0.45\textwidth}
            \includegraphics[width=1\textwidth]{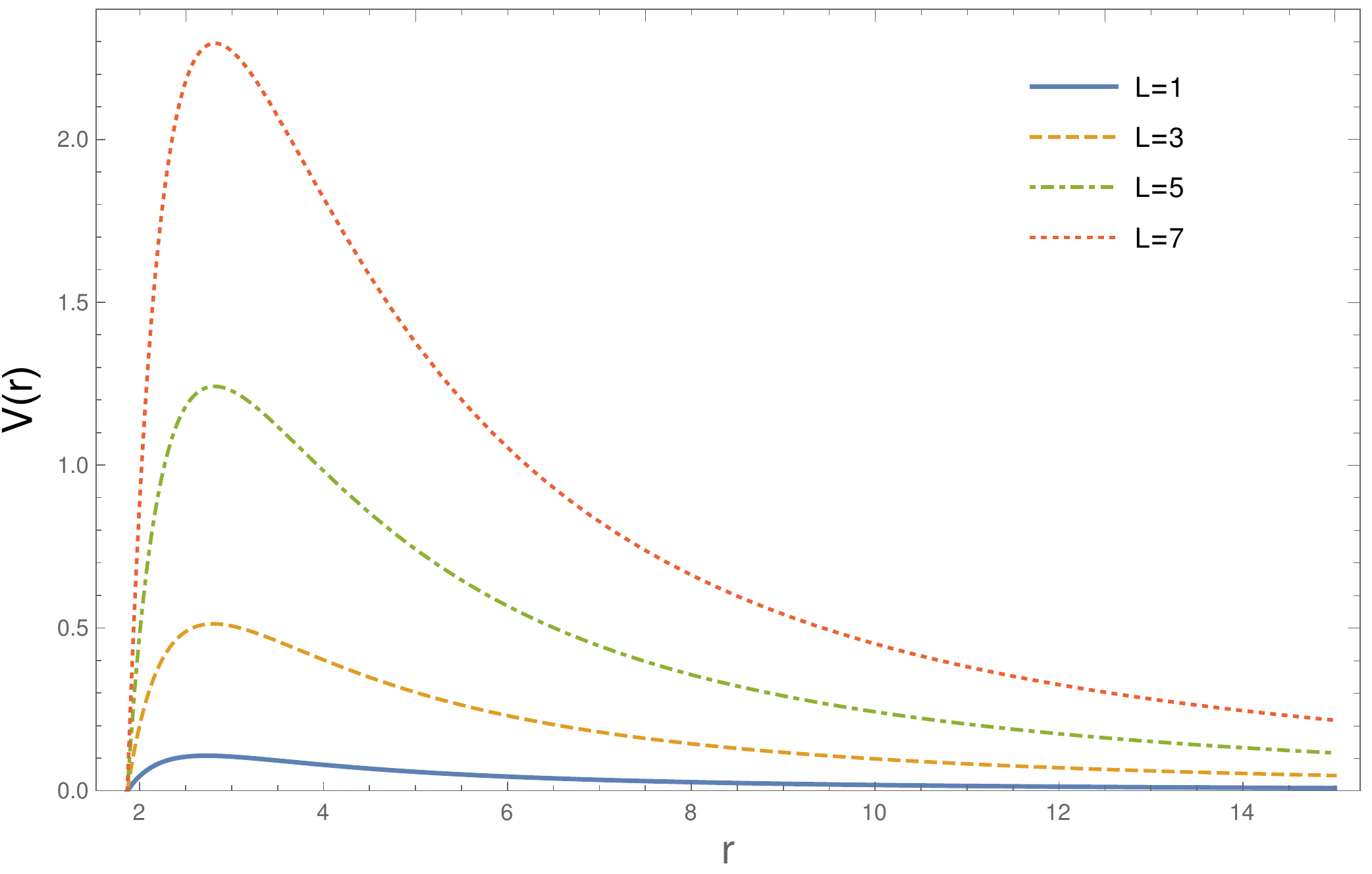}
        \end{minipage}
        \caption{Upper Left: Effective potential $V(r)$ is shown as function of $r$ keeping fixed parameters at $M=1$, $L=1$. Upper Right Effective potential $V(r)$
        is shown as function of $r$ keeping fixed parameters at $Q=0.5$, $L=1$ and Lower Centre: Effective potential $V(r)$ is shown as function of $r$ keeping
        fixed parameters at $M=1$, $Q=0.5$. }
\label{fig:a}
\end{figure*}
\section{Formation of Regular BH for Non-Linear electrodynamics}\label{Formation}
The coupling of Non-linear electrodynamics(NLED) with gravity gives rise to a regular BH for spherically symmetric space-time. Unlike singular BH, the curvature of the space-time does not blow up at $r=0$. The theory of regular BH was first proposed by Bardeen \cite{Bardeen}. Since then different types of regular black holes have been proposed \cite{Mars_1996,Barrab,Borde,AyonBeato:1998ub,Ayon-Beato1999,Ayon-Beato:2000mjt,PhysRevLett.96.031103} . The importance of the study of regular BH is that it would give an understanding of the final state of the gravitational collapse.
\par The action for gravity couple with NLED is written as
\begin{equation}\label{eq:1A}
S(g_{\mu,\nu},A_{\mu})=\int \left[\frac{R}{16\pi}-\frac{1}{4 \pi}\mathcal{L}(F)\right]\sqrt{-g}d^{4}x
\end{equation}
Where $R$ is the scalar curvature and $F$ is the electromagnetic field scalar defined as $F=\frac{1}{4}F_{\mu \nu}F^{\mu\nu}$,where $F_{\mu \nu}$ is the electromagnetic strength. The action is to be varied with respect to $g_{\mu\nu}$ in order to get Einstein's equations:
\begin{equation}\label{eq:2AB}
R_{\mu\nu}-\frac{1}{2}g_{\mu\nu} R=8\pi T_{\mu\nu}
\end{equation}
where $R_{\mu\nu}$ is the Ricci tensor and $T_{\mu\nu}$ is energy-momentum tensor defined as
\begin{equation}\label{eq:2A}
T_{\mu\nu}=\mathcal{L}(F)g_{\mu\nu}-\dfrac{d\mathcal{L}(F)}{d F} F_{\mu\sigma}F_{\nu}^{\sigma}
\end{equation}
\par In order to solve the Einstein's equation coupled with NLED, we here assume \cite{AyonBeato:1999rg,Berej:2006cc}
\begin{equation}\label{eq:3A}
\mathcal{L}(F)=F\left[1-\tanh^{2}\left(\frac{|Q|}{2 M}(2 Q^{2}F)^{\frac{1}{4}}\right)\right]
\end{equation}
where $Q$ and $M$ are the charge and mass of the BH respectively. The solution of Einstein's equation then gives rise to static spherically symmetric metric described as
\begin{equation}\label{eq:4A}
ds^{2}=-f(r) dt^{2}+\frac{1}{f(r)}dr^{2}+r^{2} d\Omega^{2}
\end{equation}
where
\begin{equation}\label{eq:5A}
f(r)=1-\frac{2 M}{r}+\frac{2 M}{r}\tanh\left(\frac{Q^{2}}{2 M r}\right)
\end{equation}
This metric is asymptotically flat, and reduces to Reissner-Nordstr{\"o}m metric for smaller value of $Q$. This metric corresponds to a charged BH,does not possess any singularity inside or on the event horizon, only if the value of charge of the source lies below $1.05M$ i.e. $\left|Q\right|\leq 1.05 M$ \cite{AyonBeato:1999rg}. The coupling of Non-linear electrodynamics (NLED) for a specific Lagrangian $\mathcal{L}(F)$ (see Eq-\eqref{eq:3A})  with gravity gives rise to  the  Ayon-Beato Garcia Regular Black Holes. Using FP duality the metric function  Eq-\eqref{eq:5A} can be reproduced within a different formulation namely by using $\mathcal{L}(F) = \frac{F}{\cosh^{2}\left( a |F/2|^{\frac{1}{4}}\right)}$, where `$a$' is a constant \cite{PhysRevD.63.044005}.
\section{Dynamical Evolution of Perturbation}\label{Dynamical}
The perturbation of BHs can be achieved in two ways: Either one can add field to the BH space-time or perturb the BH metric itself. The later one means that massive celestial objects disturb the BH space-time i.e. gravitational perturbation. In this paper, we add scalar field to the BH background and see how it couples with gravity of the BH. The Lagrangian of the minimal coupling of a perturbation of a scalar field with gravity is given by
\begin{equation}\label{eq:1}
 L_{MC}=\frac{1}{2}\nabla_{\nu}\Phi\nabla^{\nu}\Phi-\frac{1}{2}\mu^{2} \Phi^{2}
\end{equation}
If this scalar field does not backreact on the BH background, then the equation of motion(EOM) of the scalar field in the BH background would be
\begin{equation}\label{eq:2}
(\nabla_{\nu}\nabla^{\nu}-\mu^{2})\Phi=0
\end{equation}
where $\nabla_{\nu} $ is the covariant derivative and $\mu$ is
the mass of the scalar field. The metric of static scalar BH is
given by
\begin{equation}\label{eq:3}
d{s}^{2}=-f(r)dt^{2}+\frac{1}{f(r)}dr^{2}+r^{2}d\Omega^{2}
\end{equation}
\par If the background of the BH is static and spherically symmetric, one can use separation of variable method to reduce the above equation into a one dimensional  Schr{\"o}dinger-like equation. For static spherical symmetric background, we can take $\Phi$ as
\begin{equation}\label{eq:4}
\Phi(t,r,\theta,\phi)=e^{-i\omega t} Y^{m}_{L}(\theta,\phi)\frac{\Psi(r)}{r}
\end{equation}
where $Y^{m}_{L}$ are the spherical harmonics. Now the  Eq-\eqref{eq:2} simplifies to
\begin{equation}\label{eq:5}
\dfrac{d^{2}\Psi}{dr_{*}^{2}}+[\omega^{2}-V(r)]\Psi=0
\end{equation}
where $r_{*}$ is the tortoise coordinate and $V(r)$ is the potential given by
\begin{equation}\label{eq:6}
V(r)=f(r)\left(\frac{L(L+1)}{r^{2}}+\frac{1}{r}\dfrac{d f}{d r}+\mu^{2}\right)
\end{equation}
where $L$ is the multipole number.
\par To solve the Eq-\eqref{eq:5}, one has to provide the boundary conditions on $\Psi$ at event horizon and spacial infinity. Since no signal come out from an event horizon and no wave coming from infinity to further excite the BH, the choice of boundary conditions such that waves are outgoing at both the boundary \cite{1972ApJ...178..347B}. Mathematically
\begin{equation}\label{eq:7}
\Psi=\begin{cases} e^{-i \omega r_{*}} & \text{if } \hspace{3mm} r_{*}\to-\infty \\
e^{+i \omega r_{*}} & \text{if } \hspace{3mm}  r_{*} \to+\infty \end{cases}
\end{equation}
 \par The behaviour of the effective potential is shown in Fig-\ref{fig:a}. The leftmost figure shows how effective potential depends on the charge of the BH $(Q)$. The height of the potential is increased as the charge of BH  is increased. The middle figure shows how effectively potential changes when the mass of the BH $(M)$ is varied. The rightmost figure shows the change in potential when the spherical harmonics no $L$ is varied. This figure clearly shows that the height of the figure is increased as the spherical harmonics no is increased.
\begin{figure*}[!ht]
    \centering
        \begin{minipage}[b]{0.45\textwidth}
            \includegraphics[width=1.0\textwidth]{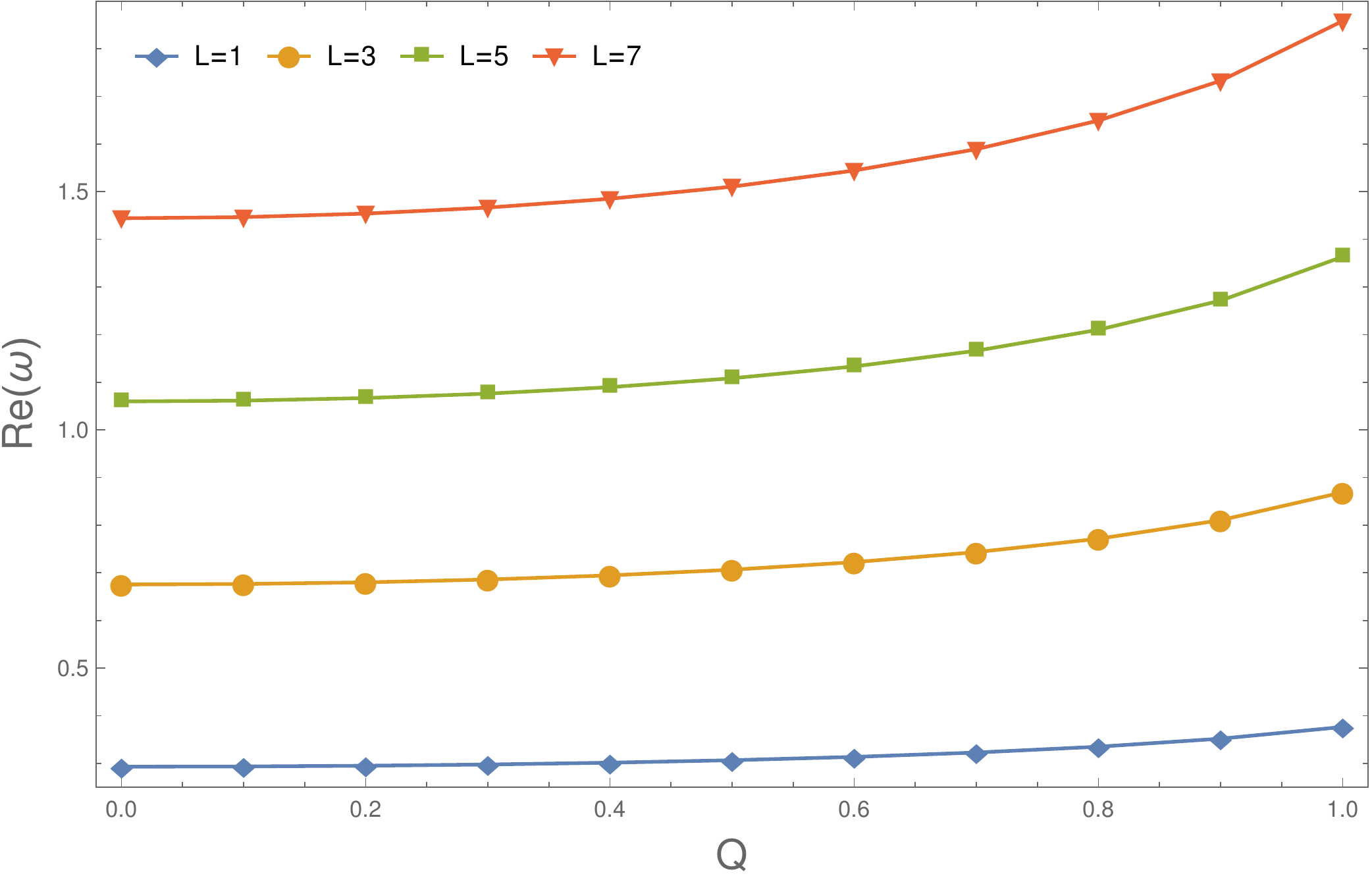}
        \end{minipage}\qquad
        \begin{minipage}[b]{0.45\textwidth}
            \includegraphics[width=1.0\textwidth]{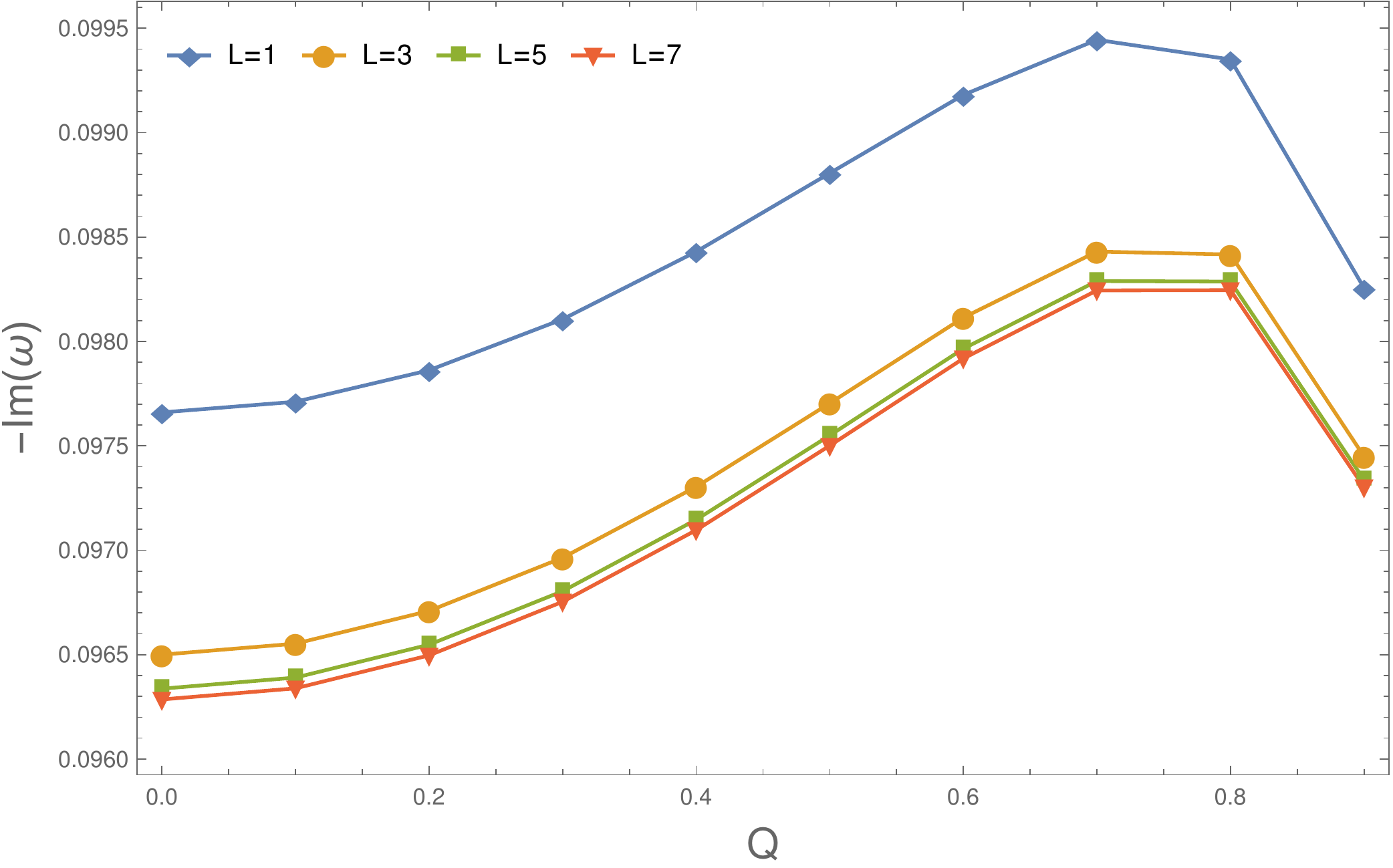}
        \end{minipage}
        \caption{Left: Real part of QNMs [$Re(\omega)$] are shown as function of Charge ($Q$) of keeping fixed parameters at $M=1$, $n=0$ and
        Right: Imaginary part of QNMs [$-Im(\omega)$] are shown as function of Charge ($Q$) of keeping fixed parameters at $M=1$,
        $n=0$}.
\label{fig:b}
\end{figure*}

\begin{figure*}[!ht]
    \centering
        \begin{minipage}[b]{0.45\textwidth}
            \includegraphics[width=1.0\textwidth]{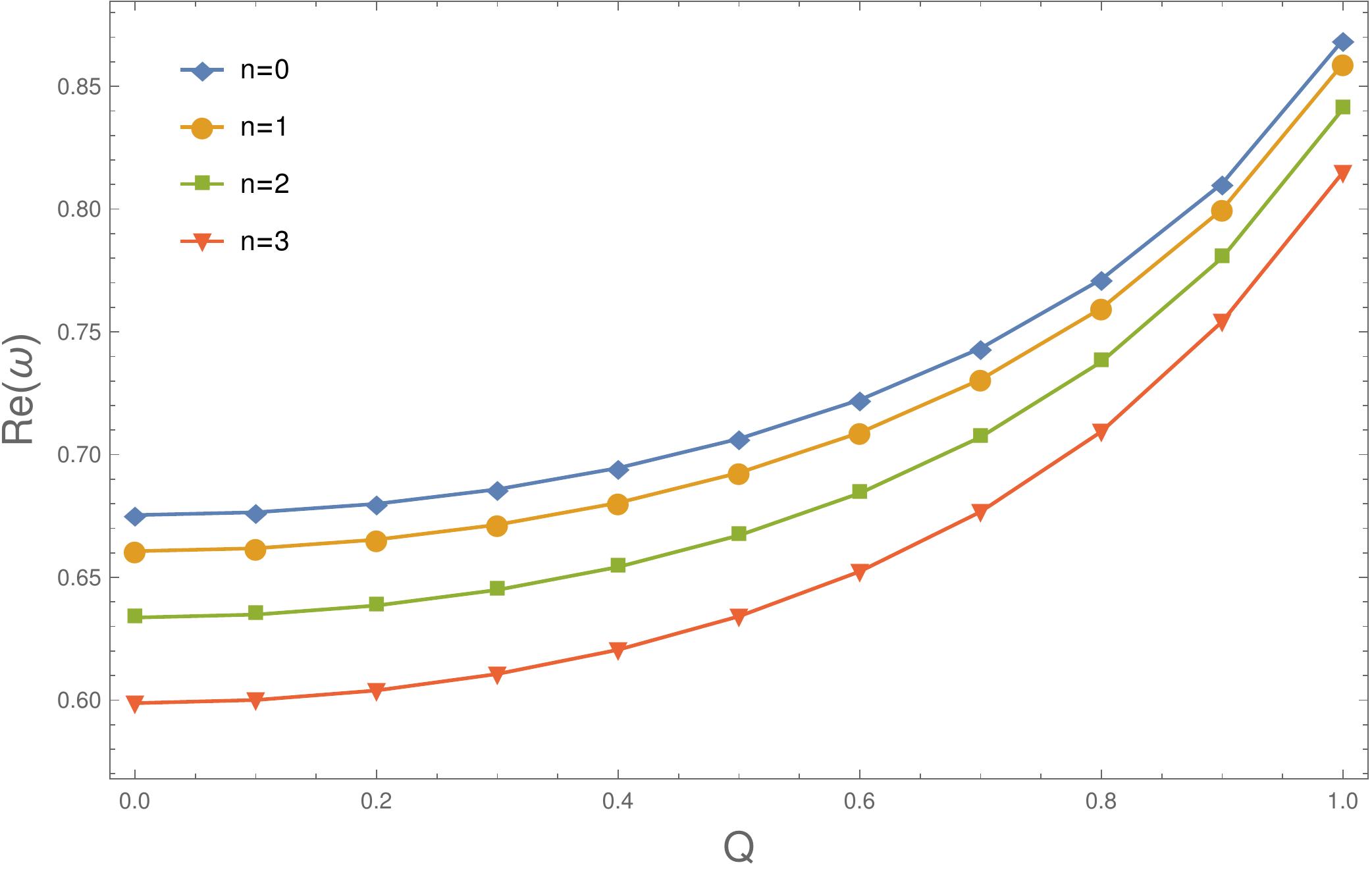}
        \end{minipage}\qquad
        \begin{minipage}[b]{0.45\textwidth}
            \includegraphics[width=1.0\textwidth]{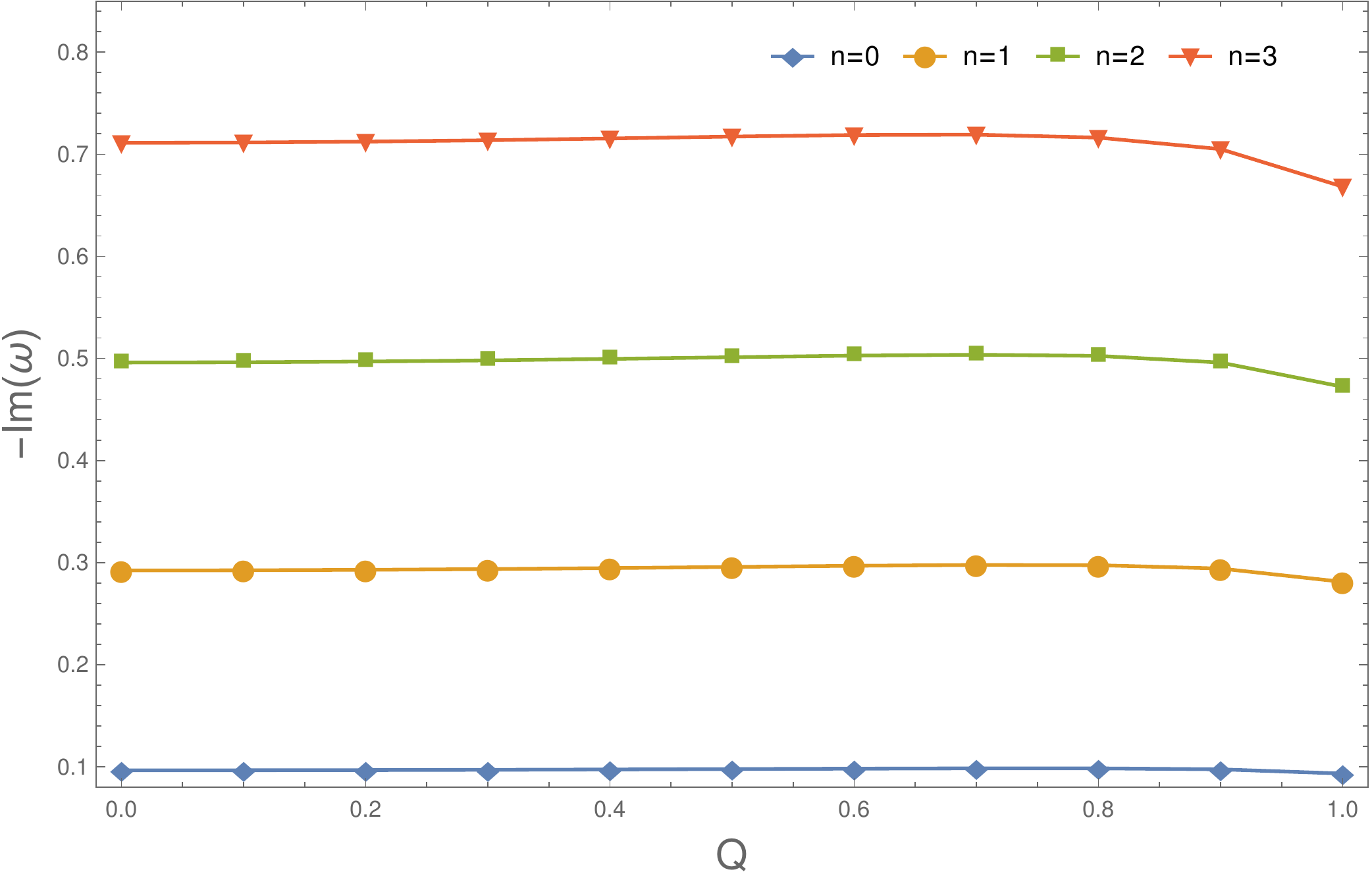}
        \end{minipage}
        \caption{Left: Real part of QNMs [$Re(\omega)$] are shown as function of Charge ($Q$) of keeping fixed parameters at $M=1$, $L=3$ and
        Right: Imaginary part of QNMs [$-Im(\omega)$] are shown as function of Charge ($Q$) of keeping fixed parameters at $M=1$, $L=3$.}
\label{fig:c}
\end{figure*}

\begin{figure*}[!ht]
    \centering
        \begin{minipage}[b]{0.45\textwidth}
            \includegraphics[width=1.0\textwidth]{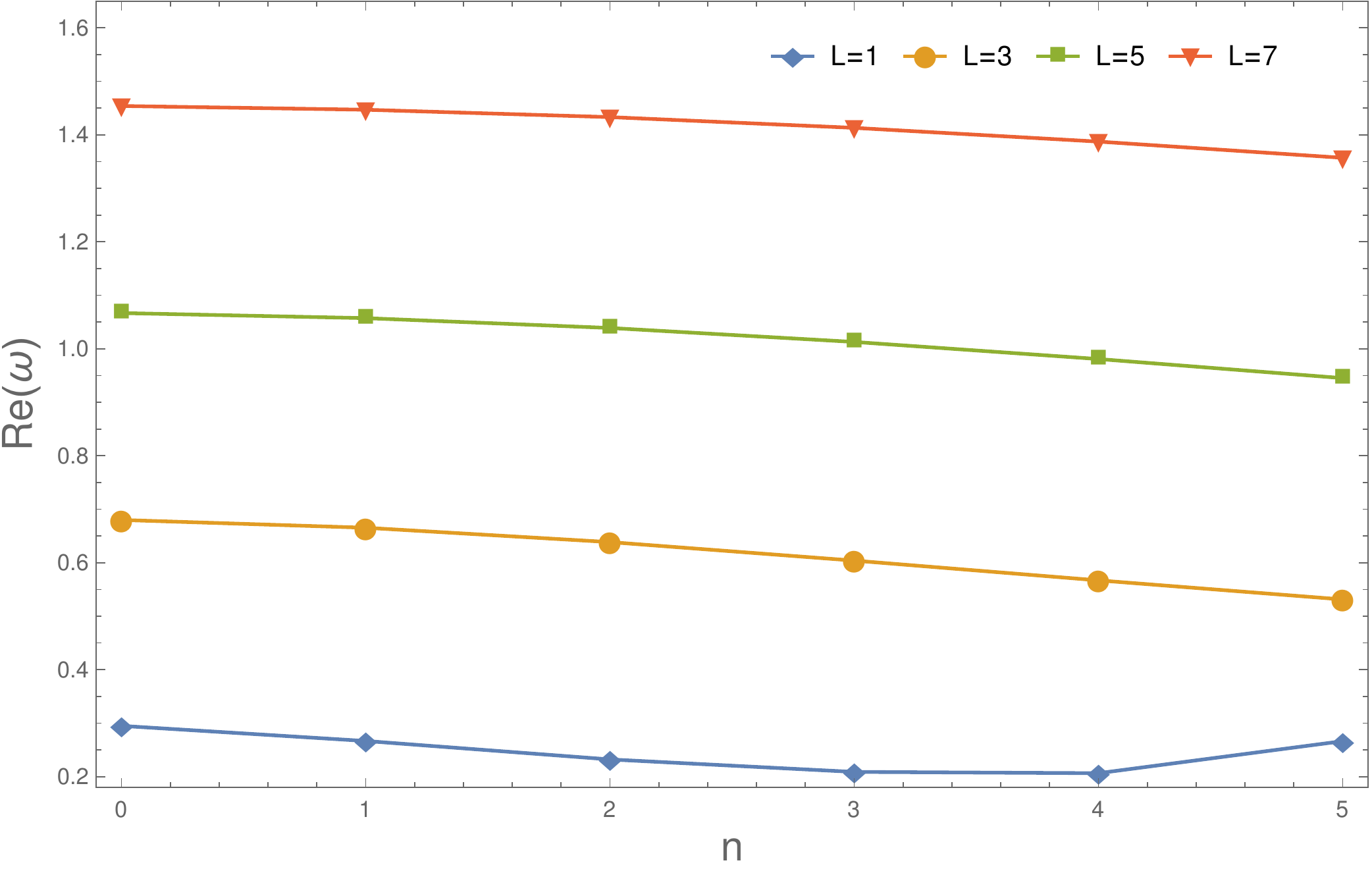}
        \end{minipage}\qquad
        \begin{minipage}[b]{0.45\textwidth}
            \includegraphics[width=1.0\textwidth]{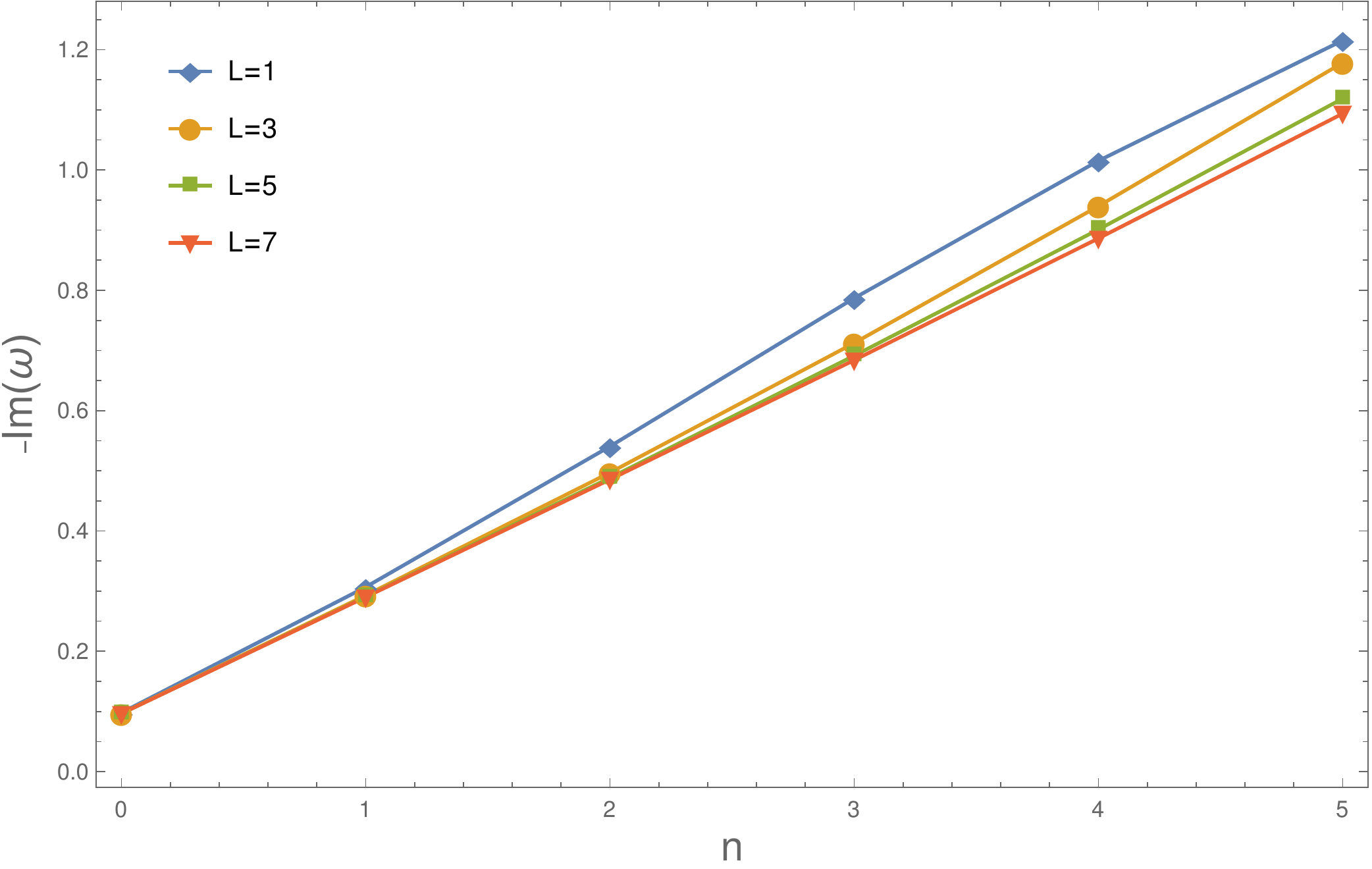}
        \end{minipage}
        \caption{Left: Real part of QNMs [$Re(\omega)$] are shown as function of mode number ($n$) of keeping fixed parameters at $M=1$, $Q=0.2$ and
        Right: Imaginary part of QNMs [$-Im(\omega)$] are shown as function of mode number ($n$) of keeping fixed parameters at $M=1$, $Q=0.2$.}
\label{fig:d}
\end{figure*}

\begin{figure*}[!ht]
    \centering
        \begin{minipage}[b]{0.45\textwidth}
            \includegraphics[width=1.0\textwidth]{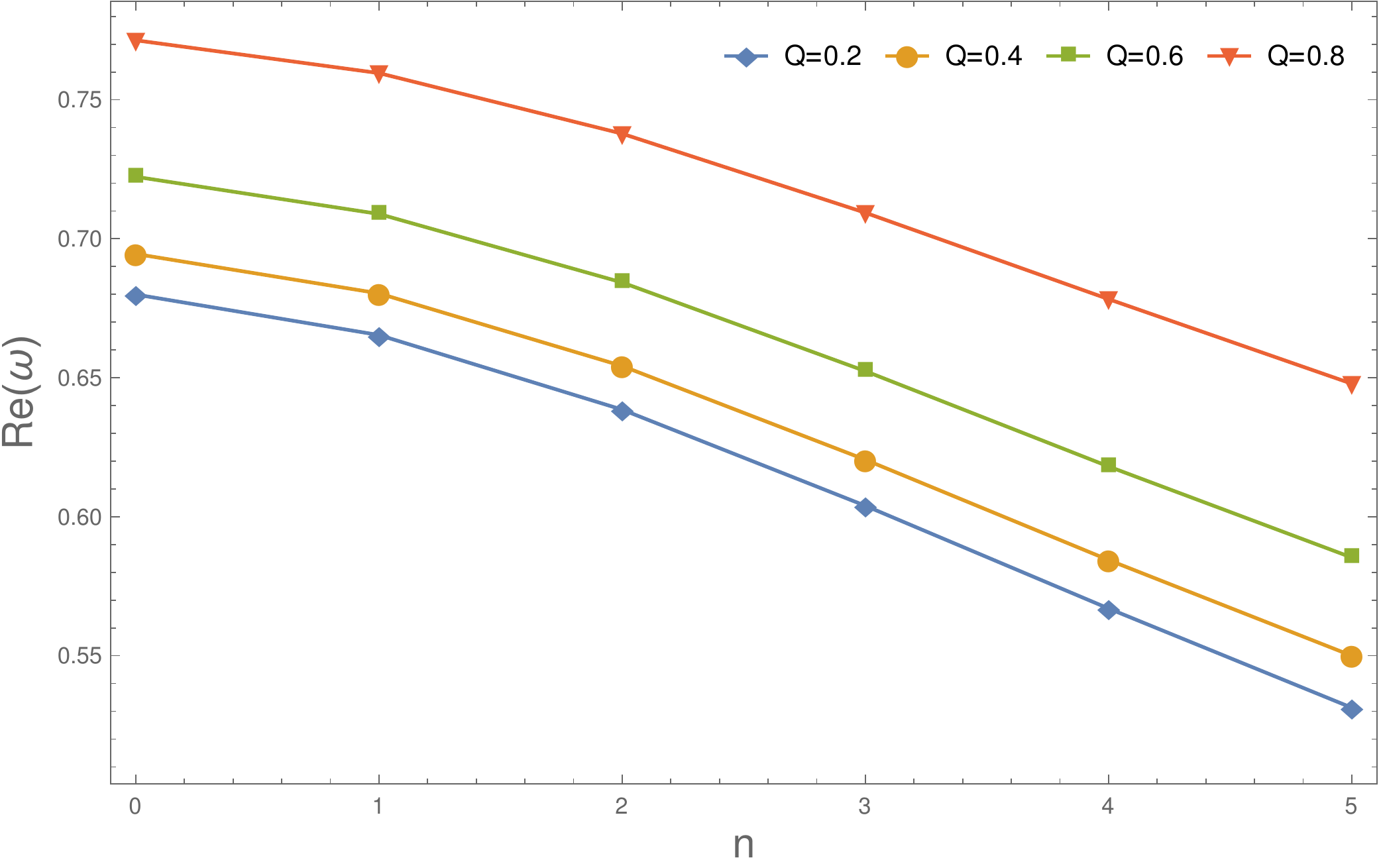}
        \end{minipage}\qquad
        \begin{minipage}[b]{0.45\textwidth}
            \includegraphics[width=1.0\textwidth]{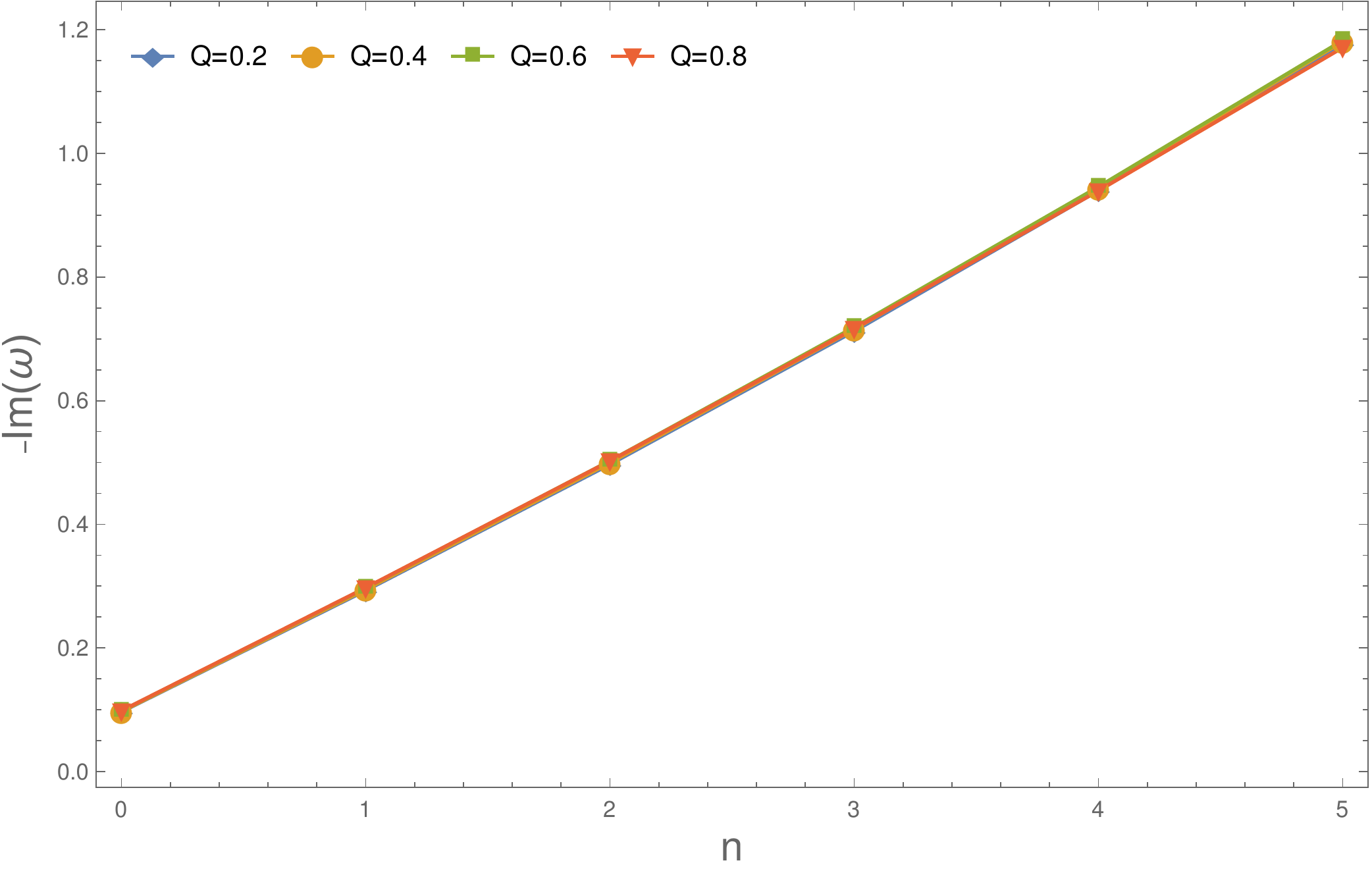}
        \end{minipage}
        \caption{Left: Real part of QNMs [$Re(\omega)$] are shown as function of mode number ($n$) of keeping fixed parameters at $M=1$, $L=3$ and
        Right: Imaginary part of QNMs [$-Im(\omega)$] are shown as as function of mode number ($n$) of keeping fixed parameters at $M=1$, $L=3$.}
\label{fig:pp}
\end{figure*}

\begin{figure*}[!ht]
    \centering
        \begin{minipage}[b]{0.45\textwidth}
            \includegraphics[width=1.0\textwidth]{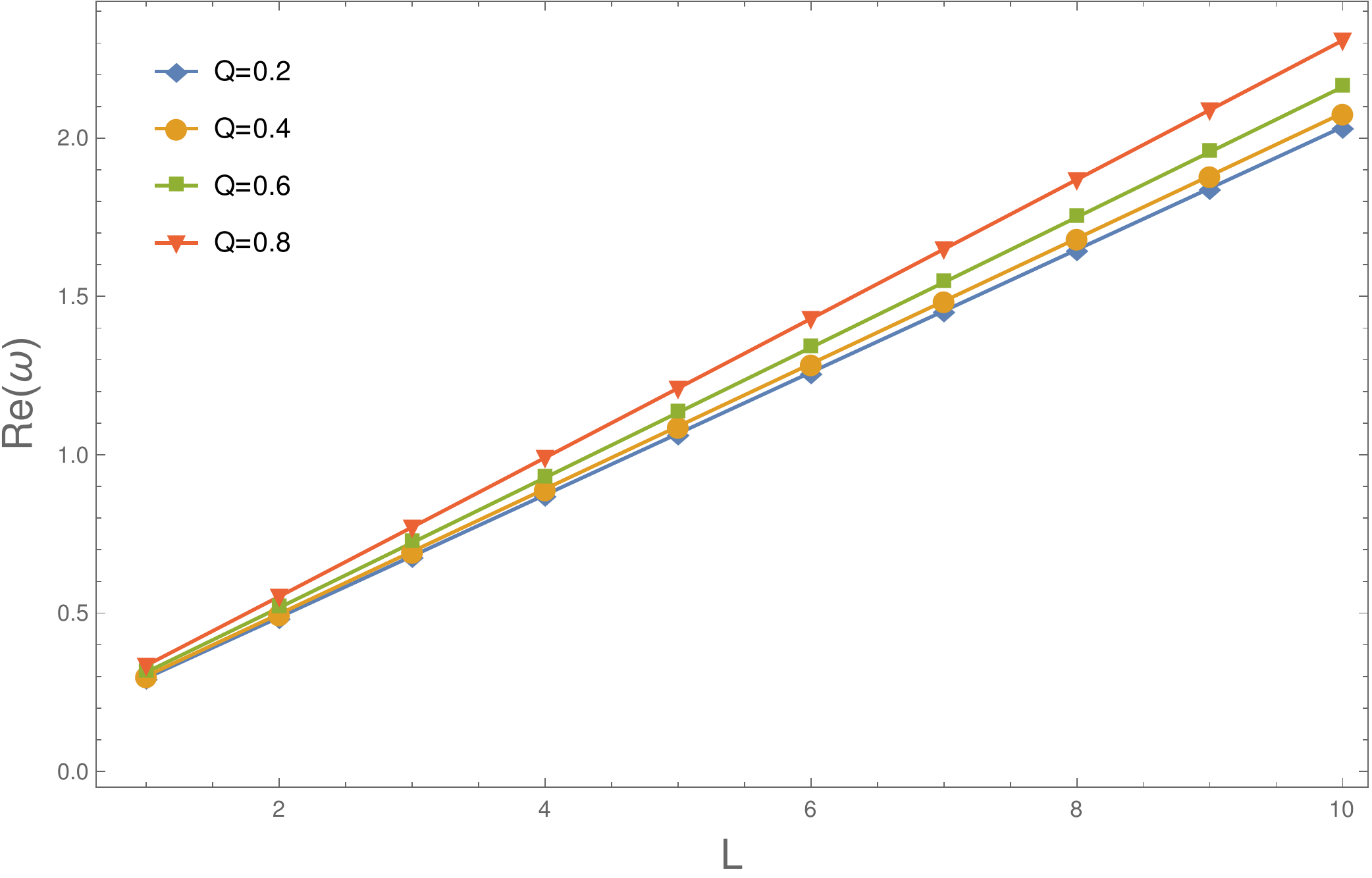}
        \end{minipage}\qquad
        \begin{minipage}[b]{0.45\textwidth}
            \includegraphics[width=1.0\textwidth]{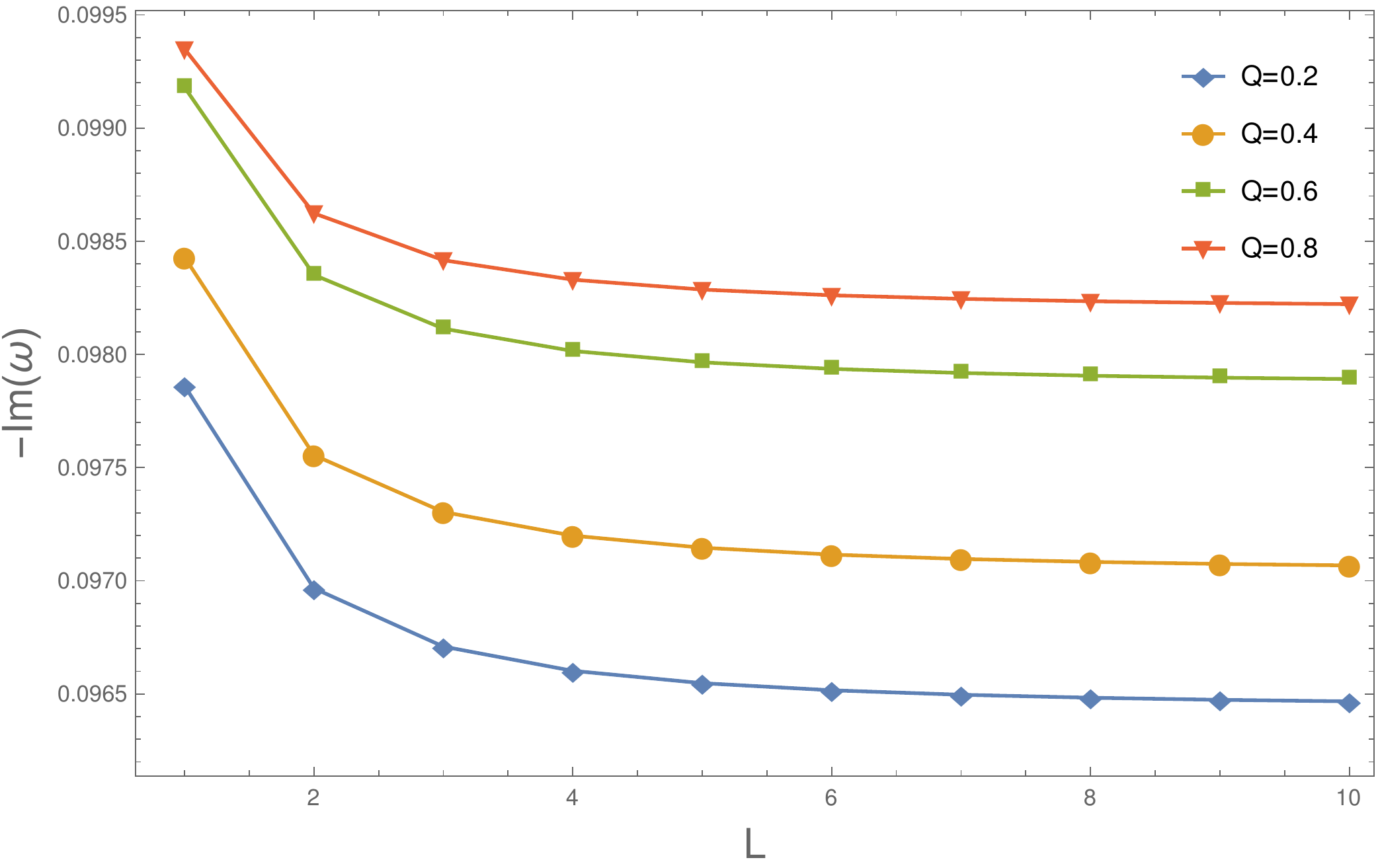}
        \end{minipage}
        \caption{Left: Real part of QNMs [$Re(\omega)$] are shown as function of multipole number $(L)$ of keeping fixed parameters at $M=1$, $n=0$ and
        Right: Imaginary part of QNMs [$-Im(\omega)$] are shown as function multipole number ($L$) of keeping fixed parameters at $M=1$, $n=0$.}
\label{fig:e}
\end{figure*}

\begin{figure*}[!ht]
    \centering
        \begin{minipage}[b]{0.45\textwidth}
            \includegraphics[width=1.0\textwidth]{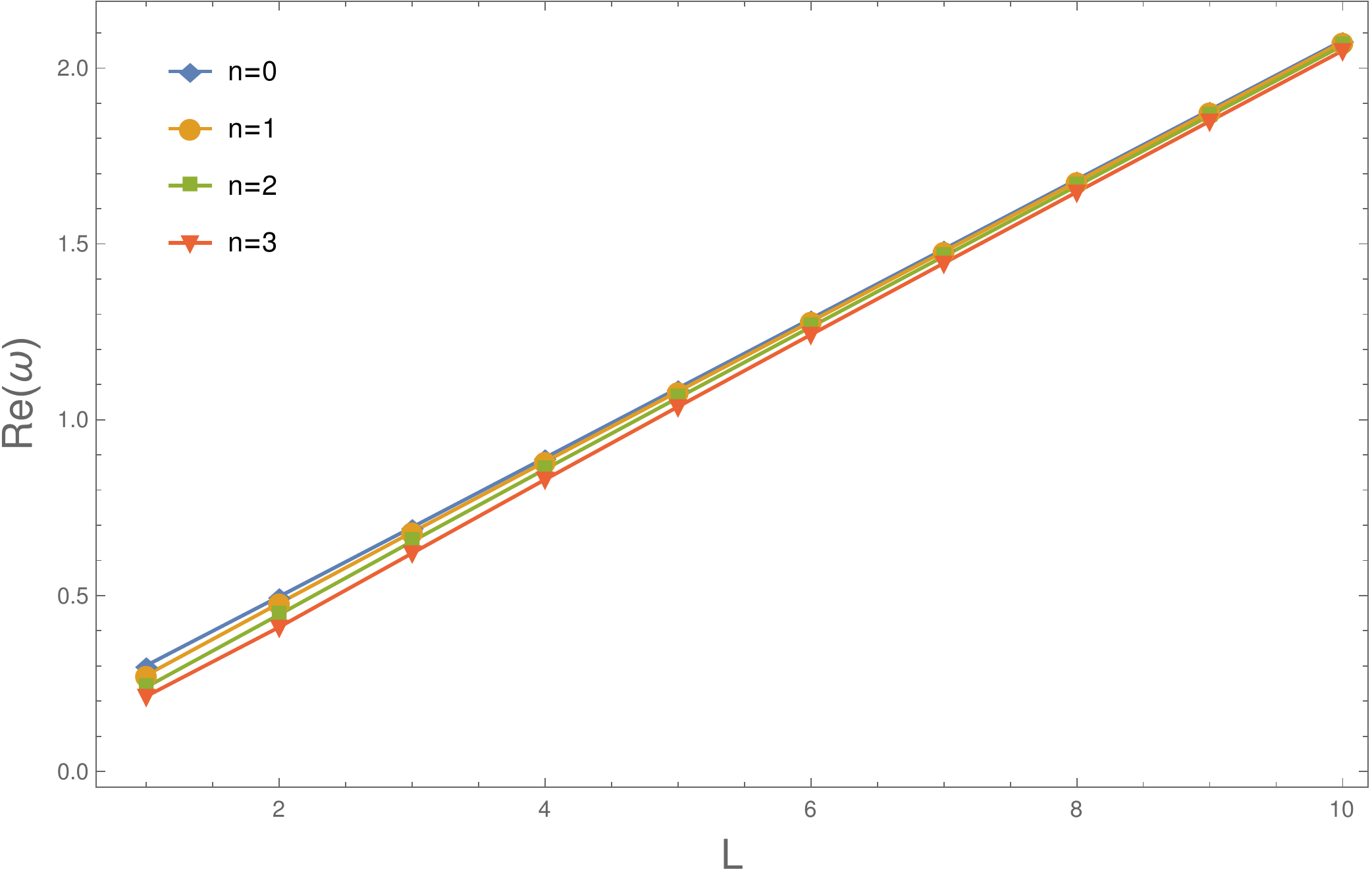}
        \end{minipage}\qquad
        \begin{minipage}[b]{0.45\textwidth}
            \includegraphics[width=1.0\textwidth]{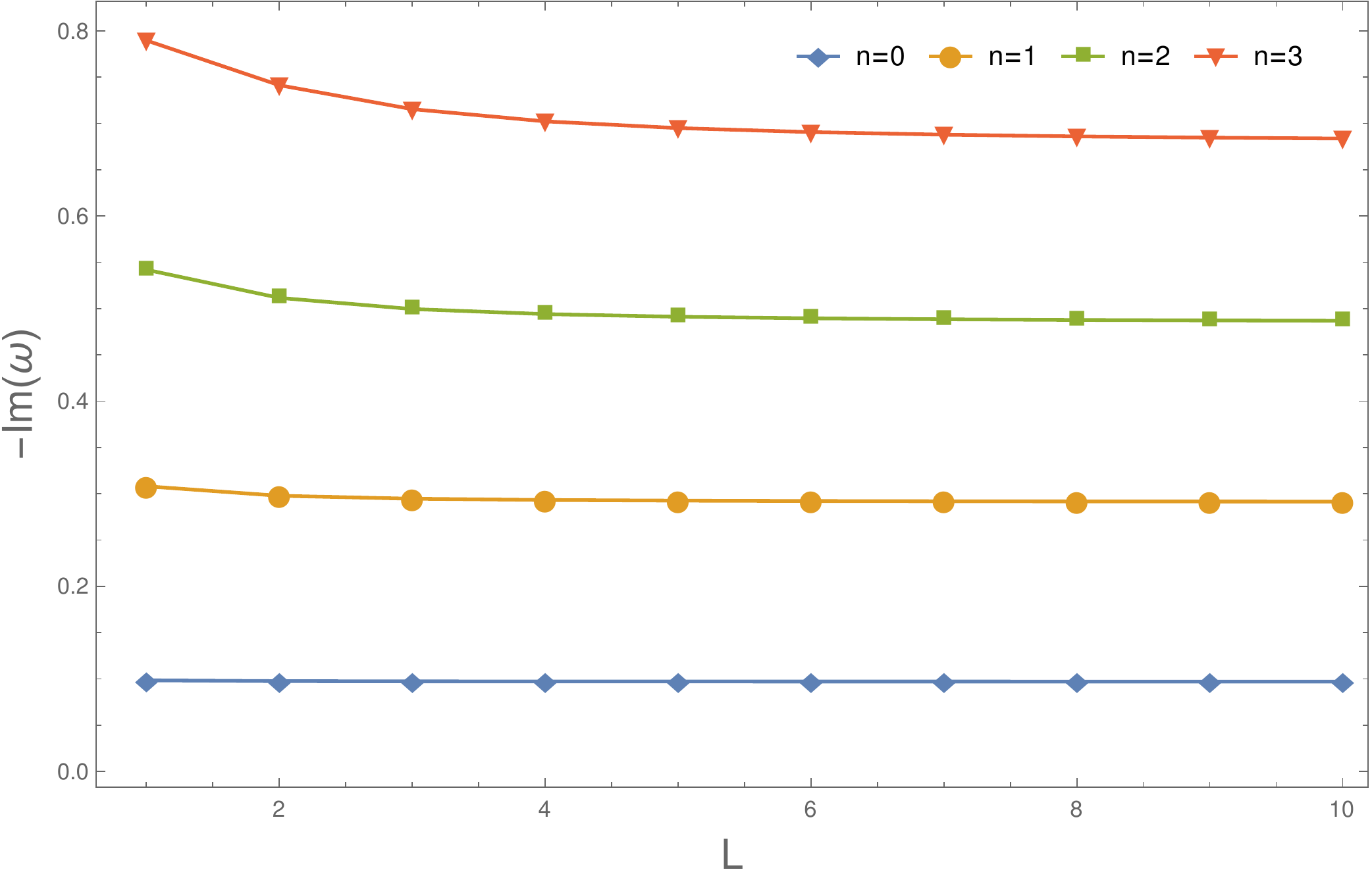}
        \end{minipage}
        \caption{Left: Real part of QNMs [$Re(\omega)$] are shown as function of multipole number $(L)$ of keeping fixed parameters at $M=1$, $Q=0.4$ and
        Right: Imaginary part of QNMs [$-Im(\omega)$] are shown as function multipole number ($L$) of keeping fixed parameters at $M=1$, $Q=0.4$.}
\label{fig:mm}
\end{figure*}

\begin{figure*}[!ht]
    \centering
        \begin{minipage}[b]{0.45\textwidth}
            \includegraphics[width=1.0\textwidth]{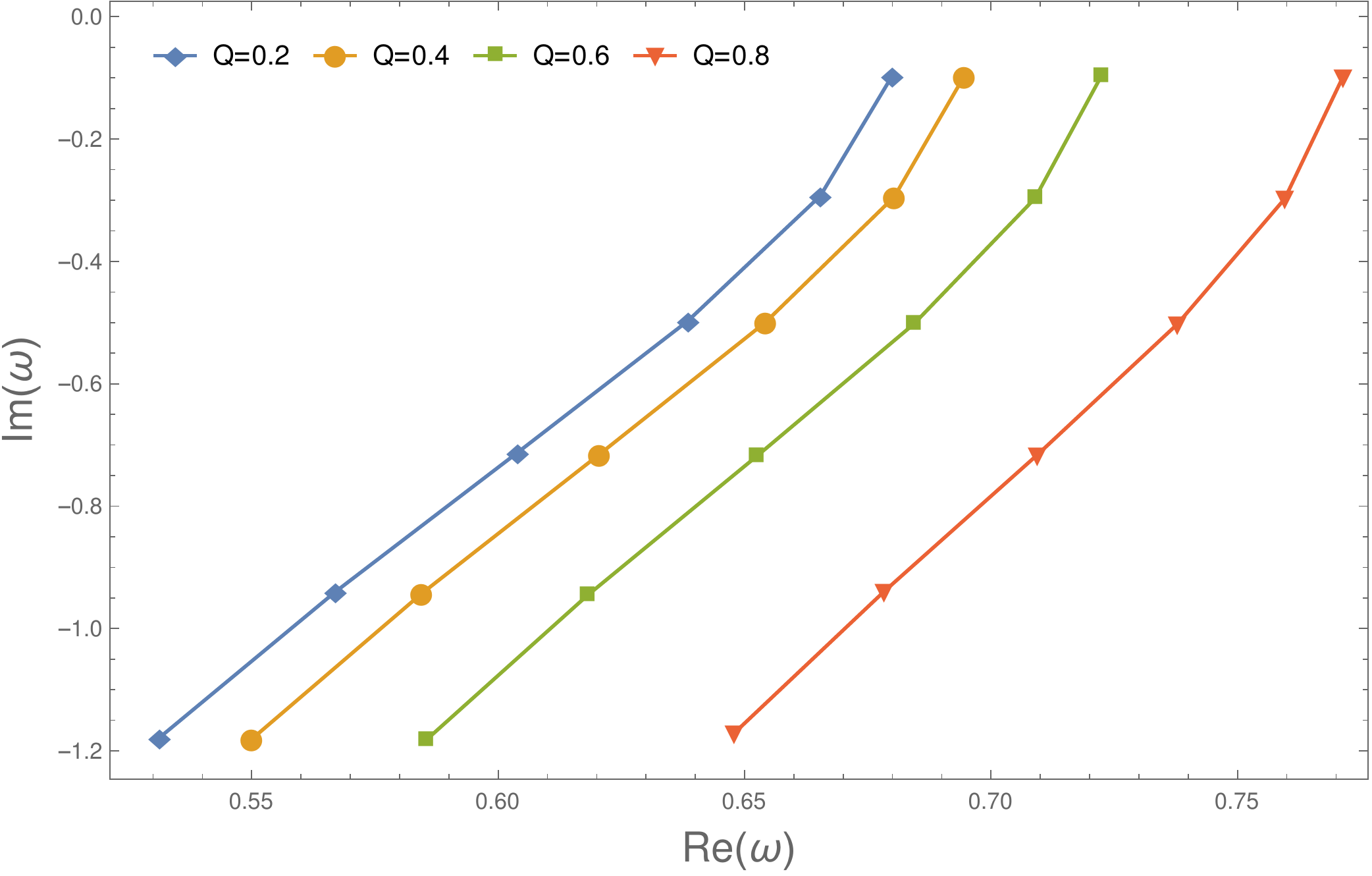}
        \end{minipage}\qquad
        \begin{minipage}[b]{0.45\textwidth}
            \includegraphics[width=1.0\textwidth]{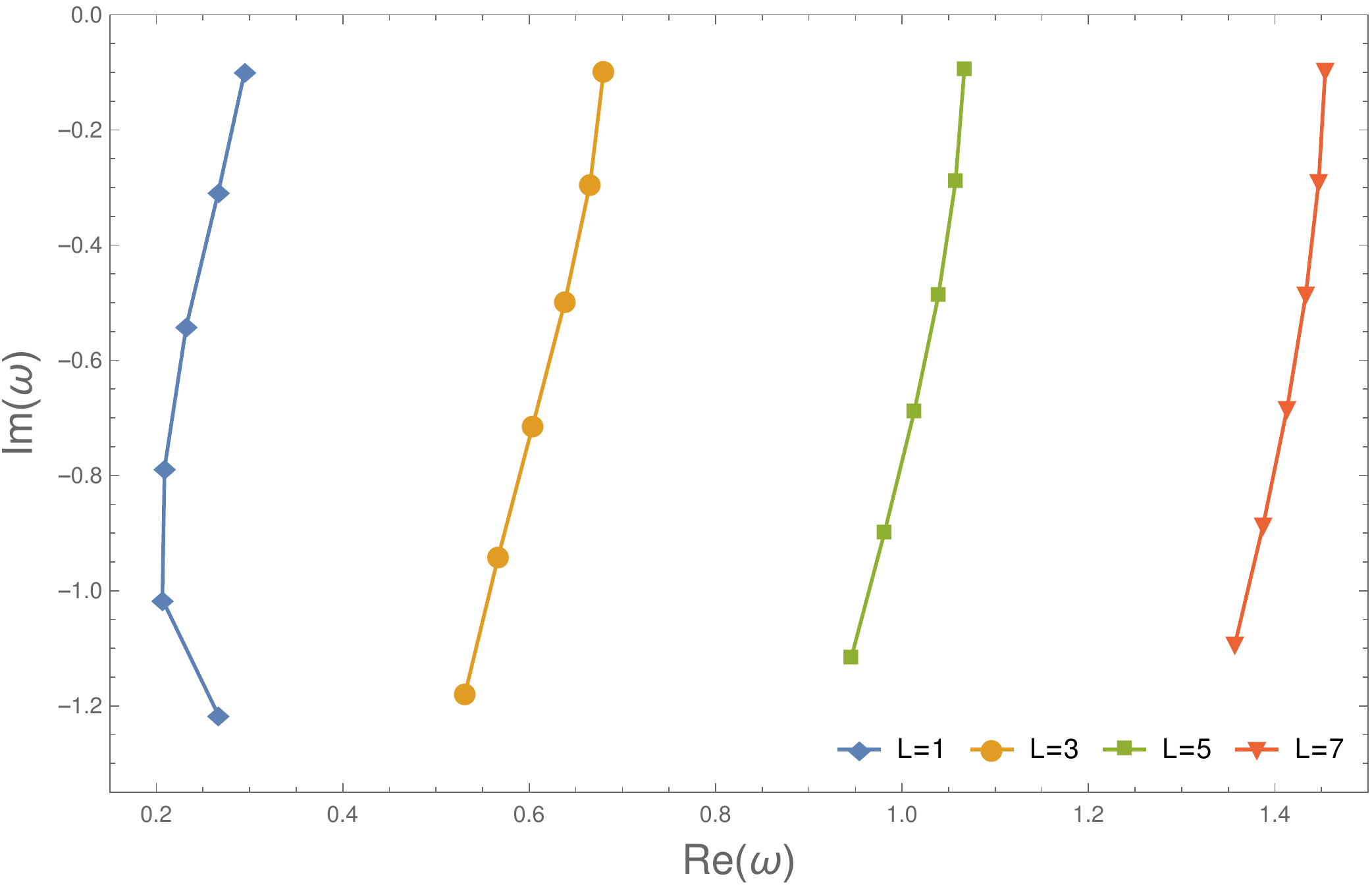}
        \end{minipage}
        \caption{Left: The QNMs are plotted by keeping fixed multiploe number at $L=3$ and Right: The QNMs are plotted by keeping fixed Charge of the BH
        at $Q=0.2$.}
\label{fig:f}
\end{figure*}

\begin{figure*}[!ht]
    \centering
        \begin{minipage}[b]{0.45\textwidth}
            \includegraphics[width=1.0\textwidth]{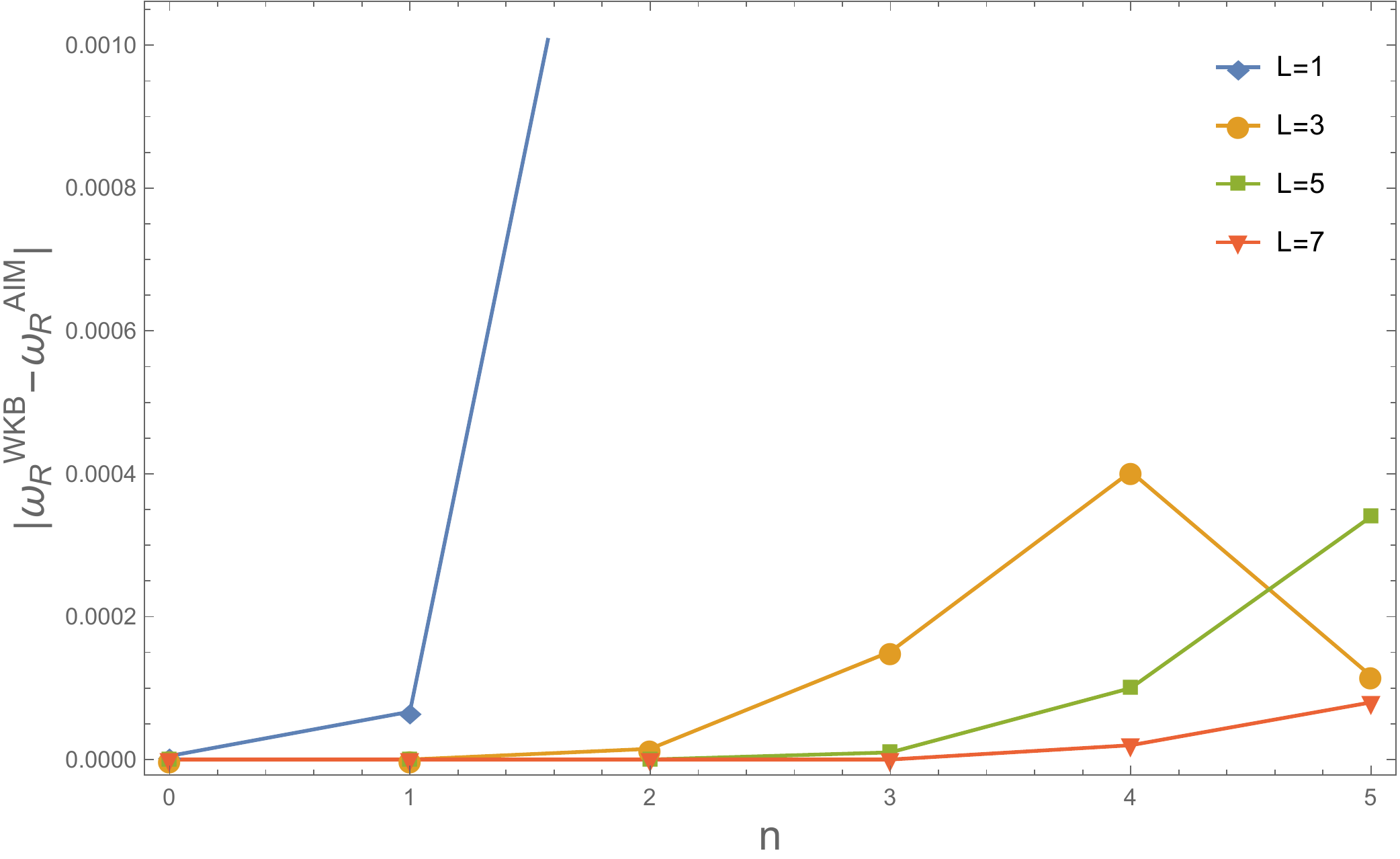}
        \end{minipage}\qquad
        \begin{minipage}[b]{0.45\textwidth}
            \includegraphics[width=1.0\textwidth]{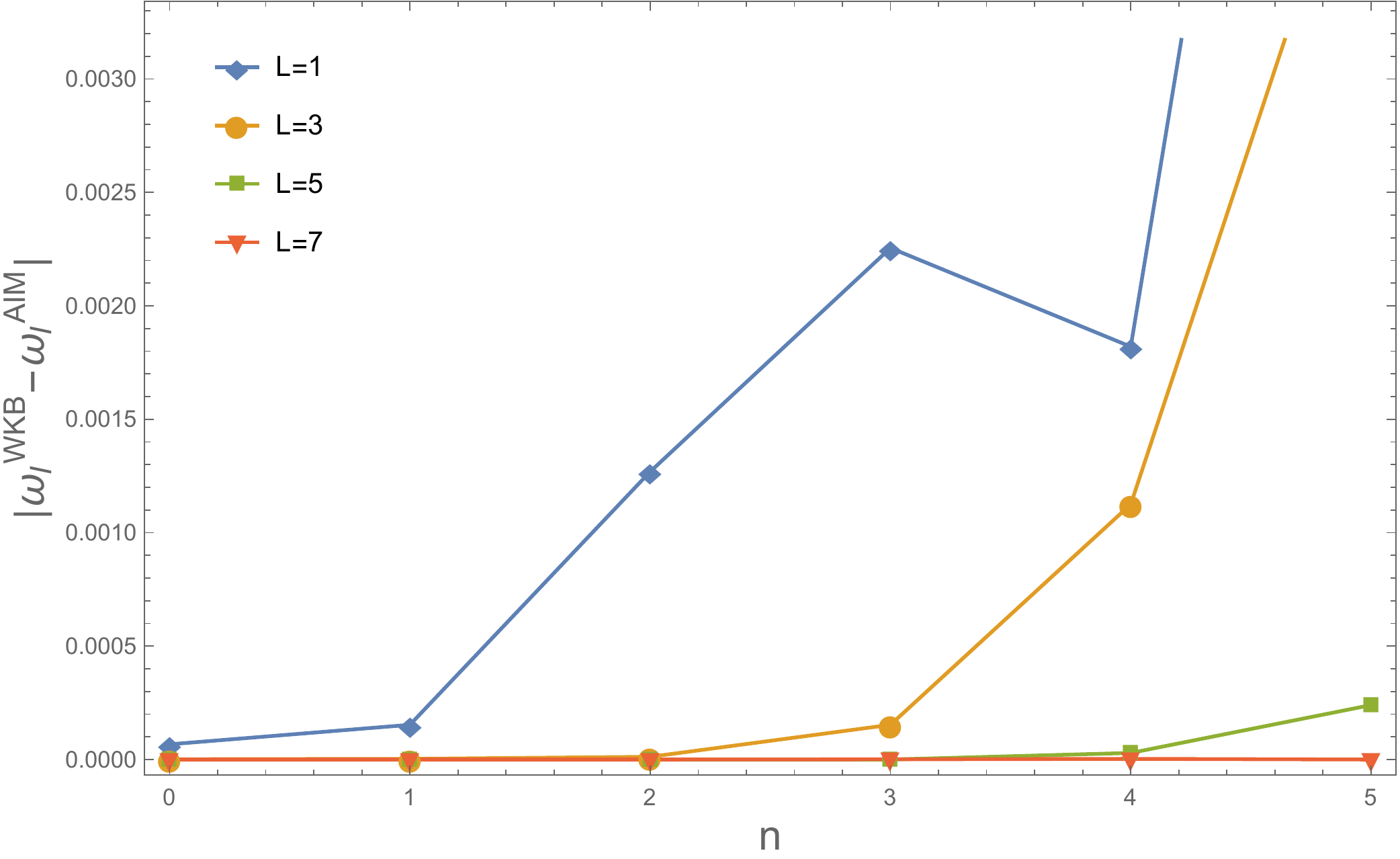}
        \end{minipage}
        \caption{ Left: The absolute difference between real part of frequency calculated using WKB and AIM $|\omega^{WKB}_{R}-\omega^{AIM}_{R}|$ vs overtone
         numbers $n$ and  Right: The absolute difference between imaginary part of frequency calculated using WKB and
         AIM $|\omega^{WKB}_{I}-\omega^{AIM}_{I}|$ vs overtone numbers $n$. }
\label{fig:g}
\end{figure*}
\begin{table}[!ht]
\centering
        \begin{tabular}{ccllll}
            \hline
            \hline
            \multirow{2}{*}{L}&\multirow{2}{*}{n}&\multicolumn{2}{c}{WKB}&\multicolumn{2}{c}{AIM}\\

             & &Re($\omega$)&Im($\omega$)&Re($\omega$)&Im($\omega$)\\
            \hline
             \hline
                         \multirow{6}{*}{1}&0        &0.294919&- 0.0979588&0.294943&- 0.0978619\\
                                           &1        &0.266699&- 0.306989 &0.266672&- 0.306742 \\
                                           &2        &0.233486&- 0.542641 &0.232042&- 0.540594\\
                                           &3        &0.224698&- 0.79528  &0.207645&- 0.791114\\
                                           &4        &0.273578&- 1.03941  &0.205633&- 1.04657\\
                                           &5        &0.431276&- 1.23769  &0.249323&- 1.28677\\

                         \multirow{6}{*}{3}&0        &0.679946&- 0.0967102&0.679946&- 0.0967092\\
                                           &1        &0.665366&- 0.292893 &0.665367&- 0.292890\\
                                           &2        &0.6385  &- 0.496942 &0.638532&- 0.496939\\
                                           &3        &0.603618&- 0.712533 &0.603940&- 0.712370\\
                                           &4        &0.565934&- 0.941304 &0.567013&- 0.939842\\
                                           &5        &0.530658&- 1.18319  &0.531917&- 1.17717\\

                         \multirow{6}{*}{5}&0        &1.06679 &- 0.0965476&1.06679&- 0.0965476\\
                                           &1        &1.05729 &- 0.290777 &1.05729&- 0.290777\\
                                           &2        &1.03889 &- 0.48835  &1.03889&- 0.488351\\
                                           &3        &1.01277 &- 0.691261 &1.01281&- 0.691269\\
                                           &4        &0.980636&- 0.901104 &0.980819&- 0.901093\\
                                           &5        &0.944502&- 1.11895  &0.945138&- 1.11868\\

                         \multirow{6}{*}{7}&0        &1.45398 &- 0.0964964&1.45398&- 0.0964964\\
                                           &1        &1.44697 &- 0.290102 &1.44697&- 0.290102\\
                                           &2        &1.43317 &- 0.485529 &1.43317&- 0.485529\\
                                           &3        &1.41307 &- 0.68393  &1.41307&- 0.683934\\
                                           &4        &1.38735 &- 0.886353  &1.38738&- 0.886365\\
                                           &5        &1.35692 &- 1.09368   &1.35705&- 1.09369\\
                         \hline
\end{tabular}
\caption{QNMs for the charge of BH $Q=0.2$ }
\label{tab1}
\end{table}
\begin{table}[!ht]
\centering
    \begin{tabular}{ccllll}
        \hline
        \hline
            \multirow{2}{*}{L}&\multirow{2}{*}{n}&\multicolumn{2}{c}{WKB}&\multicolumn{2}{c}{AIM}\\

         & &Re($\omega$)&Im($\omega$)&Re($\omega$)&Im($\omega$)\\
        \hline
        \hline
                            \multirow{6}{*}{1}&0        &0.301323&- 0.0985138&0.301339&- 0.0984298\\
                                              &1        &0.273836&- 0.308251 &0.273795&- 0.308049 \\
                                              &2        &0.241438&- 0.543728 &0.239921&- 0.542018\\
                                              &3        &0.233044&- 0.795056  &0.214119&- 0.790638\\
                                              &4        &0.281659&- 1.03594  &0.160085&-1.01101\\
                                              &5        &0.43885&- 1.22736  &0.211545&- 1.11220\\

                            \multirow{6}{*}{3}&0        &0.694544&- 0.0973048&0.694544&- 0.0973038\\
                                              &1        &0.680351&- 0.294599 &0.680351&- 0.294596\\
                                              &2        &0.654201  &- 0.499524 &0.654227&- 0.499518\\
                                              &3        &0.620254&- 0.715626 &0.620513&- 0.715466\\
                                              &4        &0.583612&- 0.944447&0.584460&- 0.943116\\
                                              &5        &0.549413&- 1.18581  &0.550303&- 1.18070\\

                            \multirow{6}{*}{5}&0        &1.08966 &- 0.0971462&1.08966&- 0.0971461\\
                                              &1        &1.08041 &- 0.292541 &1.08041&- 0.292541\\
                                              &2        &1.0625&- 0.491183  &1.06250&- 0.491184\\
                                              &3        &1.03708 &- 0.695004 &1.03711&- 0.695010\\
                                              &4        &1.00581&- 0.905542 &1.00596&- 0.905524\\
                                              &5        &0.970648&- 1.12382  &0.971179&- 1.12357\\

                            \multirow{6}{*}{7}&0        &1.48514 &- 0.0970962&1.48514&- 0.0970962\\
                                              &1        &1.47831 &- 0.291884 &1.47831&- 0.291884\\
                                              &2        &1.46488&- 0.488442 &1.46488&- 0.488443\\
                                              &3        &1.44531 &-0.687889  &1.44531&- 0.687892\\
                                              &4        &1.42028&-0.891236 &1.42031&- 0.891245\\
                                              &5        &1.39067 &- 1.09933  &1.39078&- 1.09934\\
        \hline
    \end{tabular}
\caption{QNMs for the charge of BH $Q=0.4$ }
    \label{tab2}
\end{table}
\begin{table}[!ht]
\centering
    \begin{tabular}{ccllll}
        \hline
        \hline
            \multirow{2}{*}{L}&\multirow{2}{*}{n}&\multicolumn{2}{c}{WKB}&\multicolumn{2}{c}{AIM}\\

         & &Re($\omega$)&Im($\omega$)&Re($\omega$)&Im($\omega$)\\
        \hline
        \hline
                             \multirow{6}{*}{1}&0        &0.313485&- 0.0992476&0.313490&- 0.0991805\\
                                               &1        &0.287529&- 0.309618 &0.287462&- 0.309465 \\
                                               &2        &0.256777&- 0.543962&0.255072&- 0.542693\\
                                               &3        &0.248881&- 0.792229  &0.229728&-0.789974\\
                                               &4        &0.295928&- 1.02734  &0.205885&-1.02916\\
                                               &5        &0.449067&- 1.20807 &0.182760&- 1.19983\\

                             \multirow{6}{*}{3}&0        &0.722304&- 0.098115&0.722304&- 0.0981139\\
                                               &1        &0.70892&- 0.296861 &0.708920&- 0.296858\\
                                               &2        &0.684259  &- 0.502743 &0.684274&-0.502731\\
                                               &3        &0.652232&- 0.719043 &0.652383&- 0.718890\\
                                               &4        &0.617682&- 0.947138&0.618085&- 0.946013\\
                                               &5        &0.585559&- 1.18668 &0.585442&-1.18236\\

                             \multirow{6}{*}{5}&0        &1.13315 &- 0.0979651&1.13315&- 0.0979650\\
                                               &1        &1.12443 &-0.29493 &1.12443&- 0.294930\\
                                               &2        &1.10755&- 0.494937  &1.10755&- 0.494937\\
                                               &3        &1.08358 &- 0.699785 &1.08359&- 0.699785\\
                                               &4        &1.05408&- 0.9109 &1.05418&- 0.910871\\
                                               &5        &1.02092&- 1.12922 &1.02126&- 1.12898\\

                             \multirow{6}{*}{7}&0        &1.5444 &-0.097918&1.54440&- 0.0979179\\
                                               &1        &1.53796 &- 0.294313 &1.53796&- 0.294313\\
                                               &2        &1.52529&- 0.492368 &1.52529&- 0.492368\\
                                               &3        &1.50684 &-0.693127  &1.50684&- 0.693128\\
                                               &4        &1.48324&-0.897532 &1.48326&- 0.897535\\
                                               &5        &1.45531 &- 1.10637  &1.45539&- 1.10637\\
        \hline
    \end{tabular}
    \caption{QNMs for the charge of BH $Q=0.6$ }
    \label{tab3}
\end{table}

\begin{table}[!ht]
\centering
    \begin{tabular}{ccllll}
        \hline
        \hline
            \multirow{2}{*}{L}&\multirow{2}{*}{n}&\multicolumn{2}{c}{WKB}&\multicolumn{2}{c}{AIM}\\

         & &Re($\omega$)&Im($\omega$)&Re($\omega$)&Im($\omega$)\\
        \hline
        \hline
                             \multirow{6}{*}{1}&0        &0.33491&- 0.0994018&0.334902&- 0.0993503\\
                                               &1        &0.311883&- 0.308354 &0.311771&- 0.308232 \\
                                               &2        &0.283847&- 0.53781&0.281917&- 0.537101\\
                                               &3        &0.275118&- 0.778527  &0.257058&-0.778120\\
                                               &4        &0.314347&- 1.00449 &0.241982&-1.02535\\
                                               &5        &0.447416&- 1.17416&0.225627&- 1.26343\\

                             \multirow{6}{*}{3}&0        &0.771405&- 0.0984174&0.771405&- 0.0984163\\
                                               &1        &0.759602&- 0.297409&0.759601&- 0.297404\\
                                               &2        &0.737804  &- 0.502491 &0.737805&-0.502472\\
                                               &3        &0.709354&- 0.716443 &0.709365&- 0.716304\\
                                               &4        &0.678459&- 0.940391&0.678290&- 0.939600\\
                                               &5        &0.649543&- 1.17386 &0.647944&-1.17118\\

                             \multirow{6}{*}{5}&0        &1.21013&- 0.0982866&1.21013&- 0.0982865\\
                                               &1        &1.20244 &-0.295747&1.20244&- 0.295747\\
                                               &2        &1.18754&- 0.495812  &1.18755&- 0.495811\\
                                               &3        &1.16637&- 0.700008 &1.16638&- 0.700002\\
                                               &4        &1.14025&- 0.909541&1.14028&- 0.909500\\
                                               &5        &1.11078&- 1.1252 &1.11086&- 1.12499\\

                             \multirow{6}{*}{7}&0        &1.64928 &-0.0982455&1.64928&- 0.0982455\\
                                               &1        &1.6436 &- 0.295216&1.64360&- 0.295216\\
                                               &2        &1.63244&- 0.49361 &1.63244&- 0.493609\\
                                               &3        &1.61616 &-0.694316 &1.61616&- 0.694315\\
                                               &4        &1.59532&-0.898132 &1.59532&- 0.898129\\
                                               &5        &1.57062 &- 1.10573  &1.57064&- 1.10571\\
        \hline
    \end{tabular}
    \caption{QNMs for the charge of BH $Q=0.8$ }
    \label{tab4}
\end{table}

\section{Methods of Calculations of QNMs}\label{Methods}
\subsection{WKB Approximation}\label{WKB}
The solution of the Schr{\"o}dinger-like equation\eqref{eq:5} can be achieved by both fully numerical and semi-analytical method based on the problem under consideration. Mashhoon \cite{Mashh} suggested that the WKB could be used to compute QNMs. WKB technique was first developed by Schutz and Will \cite{Schutz:1985} to solve a problem of scattering around BHs. Iyer and Will \cite{Iyer-I} in their paper extended this method to third order and Konoplya \cite{PhysRevD.68.124017, Konoplya:2004ip} extended this to sixth order. This method allows us to study the QNMs of BH systemically and is applied to any BHs problem with boundary conditions given by Eq-\eqref{eq:7}. In this method, we match two asymptotic wave function near the position of maximized potential. The QNMs are to be calculated using the WKB method of sixth-order by a formula given by
\begin{equation}\label{eq:8}
\frac{i (\omega^{2}-V_{o})}{\sqrt{-2 V^{''}_{o}}}-\sum_{j=2}^{6}\Lambda_{j}=n+\frac{1}{2}
\end{equation}
where $V_{o}$ is the maximized potential and $V^{''}_{o}$ is the second-order potential w.r.t tortoise coordinate at the position of maximized potential and $\Lambda_{j}$ is the jth order correction term of the WKB approximation. These correction terms $\Lambda_{2}$,$\Lambda_{3}$ and $\Lambda_{4}$,$\Lambda_{5}$,$\Lambda_{6}$ can be found in \cite{Iyer-II} and \cite{PhysRevD.68.024018} respectively.
\subsection{The Asymptotic Iteration Method(AIM)}\label{AIM}
First we consider a second order homogeneous differential equation for the function $\Psi$ as
\begin{equation}\label{eq:9}
\dfrac{d^{2}\Psi(x)}{d x^{2}}=\lambda_{o}(x)\dfrac{d \Psi(x)}{d x}+s_{o}(x) \Psi(x)
\end{equation}
where $\lambda_{o}(x)$ and $s_{o}(x)$ are the continuous function in between the points a and b. The above equation can be taken  derivative upto $(n+1)$th and $(n+2)$th order, where $n=1,2,3,...$. Therefore we have
\begin{subequations}\label{eq:10}
\begin{align}
\dfrac{d^{n+1}\Psi(x)}{d x^{n+1}} &=\lambda_{n-1}(x)\dfrac{d \Psi(x)}{d x}+s_{n-1}(x) \Psi(x)\\
\dfrac{d^{n+2}\Psi(x)}{d x^{n+2}} &=\lambda_{n}(x)\dfrac{d \Psi(x)}{d x}+s_{n}(x) \Psi(x)
\end{align}
\end{subequations}
where
\begin{subequations}\label{eq:11}
\begin{align}
\lambda_{n}(x)&=\frac{d \lambda_{n-1}(x)}{d x}+s_{n-1}(x)+\lambda_{0}(x)\lambda_{n-1}(x) \\
s_{n}(x) &=\dfrac{d s_{n-1}(x)}{d x}+s_{o}(x)\lambda_{n-1}(x)
\end{align}
\end{subequations}
these two recurrence relations was first found by Ciftci \cite{Ciftci_2003,article}.
\par We introduce the asymptotic aspect of the method by choosing the sufficient large value of n. Therefore we have
\begin{equation}\label{eq:12}
\frac{s_{n}(x)}{\lambda_{n}(x)}=\frac{s_{n-1}(x)}{\lambda_{n-1}(x)} \equiv \alpha(x)
\end{equation}
and the quantization condition is given by
\begin{equation}\label{eq:13}
\Delta_{n}=s_{n} \lambda_{n-1}-s_{n-1}\lambda_{n}  \hspace{3mm}  \text{n=1,2,3,..}
\end{equation}
The solution of the equation then can be written as
\begin{multline}
\Psi(x)=\exp{\left(\int^{x}\alpha(y) dy\right)} *\\
\left[ C_{2}+ 
 C_{1}\int^{x}\exp{\left\{{\int^{y} \left( \lambda_{o}(z)+2\alpha(z) \right) dz}\right\}dy} \right]
\end{multline}
where $C_{1}$ and $C_{2}$ are two constants.

\subsubsection{The Improved AIM}\label{Improved}
Since the above two recurrence relations involves derivative of the previous iteration, it takes lots of time to calculate these recurrence relations, time increases exponentially as no of iteration increase. In the following improved version of AIM, all the derivatives are excluded at each step to make this method time-efficient \cite{Cho_2010}. To do the aforesaid, $\lambda_{n}$ and $s_{n}$ are expanded in a Taylor series around the point at which one wants to perform the AIM, $x_{m}$

\begin{subequations}\label{eq:15}
\begin{align}
\lambda_{n}(x_{m})&=\sum_{i=o}^{\infty} c_{n}^{i}(x-x_{m})^{i} \\
s_{n}(x_{m}) &=\sum_{i=0}^{\infty} d_{n}^{i}(x-x_{m})^{i}
\end{align}
\end{subequations}
where $c_{n}^{i}$ and $d_{n}^{i}$ are the ith Taylor coefficient of $\lambda_{n}(x_{m}$ and $s_{n}(x_{m})$ respectively. Now we substitute the above expression into the previous set of recursion relation to get new set of recursion relation for the coefficient
 \begin{subequations}\label{eq:16}
\begin{align}
c_{n}^{i} &=(i+1) c_{n+1}^{i+1}+d_{n-1}^{i}+\sum_{k=0}^{i} c_{0}^{k}c_{n-1}^{i-k} \\
d_{n}^{i} &=(i+1) d_{n-1}^{i+1}+\sum_{k=0}^{i} d_{0}^{k}c_{n-1}^{i-k}
\end{align}
\end{subequations}
 These recursion relations do not involve any derivative operators. The quantization condition for this improved method is given by
\begin{equation}\label{eq:17}
d_{n}^{0}c_{n-1}^{0}-d_{n-1}^{0}c_{n}^{0}=0
\end{equation}
The reason behind of the choosing such quantization was explained in \cite{Cho_2010}

\section{Calculation of the QNMS using AIM}\label{cal}
We first do a coordinate transformation to make the radial equation relevant to AIM which leads to a simpler solution. For the ABG BH, it is convenient to change coordinate $r$ to $x$ as $x=\frac{1}{r}$. The radial equation then can be written in terms of $x$ as
\begin{equation}\label{eq:18}
\dfrac{d^{2}\Psi}{d x^{2}}+\dfrac{d [ln(x^{2}f)]}{d x}\dfrac{d \Psi}{d x}+\left(\frac{\omega^{2}}{x^{4}f^{2}}-\frac{L(L+1) x^{2}}{f}+\frac{1}{x f}\dfrac{d f}{d x}\right)\Psi=0
\end{equation}
Since only the outgoing waves survive at the boundary and there is a regular singularity at the event horizon,we define $\Psi$ as
\begin{equation}\label{eq:19}
\Psi=(x-x_{o})^{-\frac{i \omega}{K_{o}}} e^{-i \omega \int \frac{d x}{x^{2}f(x)}} u(x)
\end{equation}
where $x_{o}$ is the position of the event horizon and $K_{o}$ is the surface gravity at horizon defined as
\begin{equation}\label{eq:20}
K_{o}=\left.\frac{1}{2}\dfrac{d f}{d r} \right |_{r \to r_{o}}=\left. -\frac{x^{2}}{2}\dfrac{d f}{d x}\right|_{x\to x_{o}} 
\end{equation}
The Eq-\eqref{eq:18} then takes form
\begin{equation}\label{eq:21}
\dfrac{d^{2} u(x)}{d x^{2}}=\lambda_{o}(x)\dfrac{d u(x)}{d x}+s_{o}(x)u(x)
\end{equation}
with
\begin{equation}
\lambda_{o}=\frac{-x^2 \left\{3 M \sinh \left(\frac{Q^2 x}{M}\right)+Q^2 x\right \} \text{sech}^2\left(\frac{Q^2 x}{2 M}\right)+6 M x^2-2 x+2 i \omega }{x^2 \left\{2 M x \tanh \left(\frac{Q^2 x}{2 M}\right)-2 M
	x+1\right\}}+\frac{2 i \omega }{(x-x_{o}) K_{o}}
\end{equation}
\begin{small}
\begin{multline}
s_{o}=\frac{1}{x^2 (x-x_{o})^2 K_{o}^2 \left(2 M x \tanh \left(\frac{Q^2 x}{2 M}\right)-2 M x+1\right)}
\left[x^2 \omega ^2 \left\{2 M x \tanh \left(\frac{Q^2 x}{2 M}\right)-2 M x+1\right \} \right. \\
\left. +i \omega  K_{o} \left\{ x^2 \left \{2 M (2 x-3 x_{o}) \tanh \left(\frac{Q^2 x}{2 M}\right)+Q^2 x (x-x_{o}) \text{sech}^2\left(\frac{Q^2 x}{2 M}\right) \right \}-4 M x^3+x^2 (6 M
x_{o}+1)-2 x (x_{o}+i \omega )+2 i x_{o} \omega \right \}\right. \\
\\
+\left.\frac{1}{2} (x-x_{o})^2 K_{o}^2 \text{sech}^2\left(\frac{Q^2 x}{2 M}\right) \left \{\left(L^2+L+2 M x\right) \cosh \left(\frac{Q^2 x}{M}\right)+L^2+L+2 x \left(M-Q^2 x\right)-2 M x \sinh
\left(\frac{Q^2 x}{M}\right)\right \}\right]
\end{multline}
\end{small}
To calculate QNMs using AIM, we first find the event horizon in term of $x$ by expanding $f(x)$ about $x=0$ and kept up to 50 terms, then we  equate with expanded $f(x)$ to zero to get the event horizon. After finding the value of $ \lambda_{o}$ and $ s_{o}$, we expand $\lambda_{o}$ and $s_{o}$ in Taylor series about the point where the effective potential get maximized. The ith components of the Taylor series of  $\lambda_{o}$ and $s_{o}$  are identified as $c_{0}^{i}$ and $d_{0}^{i}$ respectively. Then using the recurrence relation Eq-\eqref{eq:16}, we find other $c_{n}^{i}$ and $d_{n}^{i}$. At some large value of n, we truncate the iteration. Then we solve the Eq-\eqref{eq:17} to find the QNMs frequency. We use Mathematica Software to perform all the calculations.

\section{conclusion}\label{conclusion}
We calculate QNMs of ABG regular BH for the scalar field
using two methods WKB and AIM. We show the calculated numerical values of QNMs in table \ref{tab1}-\ref{tab4}. The calculation shows that the
difference between WKB and AIM significantly reduces as the
multipole number increases (see Fig-\ref{fig:g}). WKB method works correctly for $n \leq
L$ .For lower multipole number and higher mode, the calculated
QNMs using AIM is more acceptable than WKB.
\par The Fig-\ref{fig:b} and Fig-\ref{fig:c} show that the real part of QNMs frequency monotonically increases but the imaginary part first increases and
gets maximized around $Q=0.7$ then decreases rapidly
(Fig-\ref{fig:b}) as the value of $Q$ increases. The
Fig-\ref{fig:d} and Fig-\ref{fig:pp} show that the real part of
frequency decreases and imaginary part linearly increases as the
mode number ($n$) increases. The Fig-\ref{fig:e} and
Fig-\ref{fig:mm} show that the real part of frequency
approximately linear with multipole number ($L$) and the
imaginary part is greatest for the lower value of multipole
number ($L$).We see the imaginary part gets independent of
multipole number ($L$) for larger multipole number ($L$) which
reflects the effect of the eikonal limit i.e. $L \to
\infty$ . Therefore, the stability of ABG BH is highest around
the charge of the BH $Q=0.7$ against the scalar field for BH mass
$M=1$. All the QNMs are found stable. How the real and imaginary part of QNMs is related to each other is shown in Fig-\ref{fig:f}
 \par The QNMs of different regular charged BH for the massless scalar field was studied in \cite{PhysRevD.87.024034}. Our result is consistent in comparison to the findings of \cite{PhysRevD.87.024034} and reproduces known results in the Schwarzschild limit.We extend the works of the authors of \cite{PhysRevD.87.024034} for a particular metric given by references [4,5] in \cite{PhysRevD.87.024034} by studying how QNMs are dependent on the BH parameters in details and also find the heights possible region of charge of BH for which BH is most stable. 
\section*{Acknowledgments}
MM is thankful to CSIR-UGC for providing financial support. FR
and MK are grateful to the Inter-University Centre for Astronomy
and Astrophysics (IUCAA), Pune, India for providing Associateship
programme under which a part of this work was carried out. FR is
also thankful to DST-SERB,  Govt. of India and RUSA 2.0, Jadavpur
University,  for financial support.
\section*{Note}
Before the acceptance of our paper, we were unaware that our work is closely related to these \cite{refId0,Li:2016oif,Lopez:2018aec,Panotopoulos:2019qjk,PhysRevD.103.124050,Cai:2020kue} works. Although, the lapse function of our ABG BH is different from the lapse function of the ABG BHs used by many authors to compute their QNMs. No authors have studied the QNMs of the ABG BH having this same lapse function in detail as we have done in our paper.


@article{Konoplya:2011qq,
      author         = "Konoplya, R. A. and Zhidenko, A.",
      title          = "{Quasinormal modes of black holes: From astrophysics to
                        string theory}",
      journal        = "Rev. Mod. Phys.",
      volume         = "83",
      year           = "2011",
      pages          = "793-836",
      doi            = "10.1103/RevModPhys.83.793",
      eprint         = "1102.4014",
      archivePrefix  = "arXiv",
      primaryClass   = "gr-qc",
      SLACcitation   = "
}
@article{Regge:1957td,
      author         = "Regge, Tullio and Wheeler, John A.",
      title          = "{Stability of a Schwarzschild singularity}",
      journal        = "Phys. Rev.",
      volume         = "108",
      year           = "1957",
      pages          = "1063-1069",
      doi            = "10.1103/PhysRev.108.1063",
      SLACcitation   = "
}

@Article{Schutz:1985, author = {Schutz, B. F. and Will, C. M.},
title = {Black hole normal modes - A semianalytic approach},
journal = {Astrophys. J.},
year = {1985},
volume = {L291},
pages ={33}
}

@Article{Mashhon:1984, author = {Hans-Joachim Blome and Bahram
Mashhoon}, title = {Quasi-normal oscillations of a schwarzschild
black hole}, journal = {Phys Lett A}, volume = {100}, pages =
{231}, year = {1984} }

@article{Leaver:1985ax,
      author         = "Leaver, E. W.",
      title          = "{An Analytic representation for the quasi normal modes of
                        Kerr black holes}",
      journal        = "Proc. Roy. Soc. Lond.",
      volume         = "A402",
      year           = "1985",
      pages          = "285-298",
      doi            = "10.1098/rspa.1985.0119",
      SLACcitation   = "
}
@article{Horowitz-Hubeny:2000,
  title = {Quasinormal modes of AdS black holes and the approach to thermal equilibrium},
  author = {Horowitz, Gary T. and Hubeny, Veronika E.},
  journal = {Phys. Rev. D},
  volume = {62},
  issue = {2},
  pages = {024027},
  numpages = {11},
  year = {2000},
  month = {Jun},
  publisher = {American Physical Society},
  doi = {10.1103/PhysRevD.62.024027},
  url = {https://link.aps.org/doi/10.1103/PhysRevD.62.024027}
}
@Article{cho:2012,
author = {H. T. Cho, A. S. Cornell, Jason Doukas, T.-R. Huang, and Wade Naylor},
title = {A New Approach to Black Hole Quasinormal Modes: A Review of the Asymptotic Iteration Method},
journal = {Advances in Mathematical Physics},
year = {2012},
OPTnumber = {281705},
OPTpages = {42}
}
@article{Zerilli:1970,
  title = {Gravitational Field of a Particle Falling in a Schwarzschild Geometry Analyzed in Tensor Harmonics},
  author = {Zerilli, Frank J.},
  journal = {Phys. Rev. D},
  volume = {2},
  issue = {10},
  pages = {2141--2160},
  numpages = {0},
  year = {1970},
  month = {Nov},
  publisher = {American Physical Society},
  doi = {10.1103/PhysRevD.2.2141},
  url = {https://link.aps.org/doi/10.1103/PhysRevD.2.2141}
}
@article{Iyer-I,
  title = {Black-hole normal modes: A WKB approach. I. Foundations and application of a higher-order WKB analysis of potential-barrier scattering},
  author = {Iyer, Sai and Will, Clifford M.},
  journal = {Phys. Rev. D},
  volume = {35},
  issue = {12},
  pages = {3621--3631},
  numpages = {0},
  year = {1987},
  month = {Jun},
  publisher = {American Physical Society},
  doi = {10.1103/PhysRevD.35.3621},
  url = {https://link.aps.org/doi/10.1103/PhysRevD.35.3621}
}
@article{Iyer-II,
  title = {Black-hole normal modes: A WKB approach. II. Schwarzschild black holes},
  author = {Iyer, Sai},
  journal = {Phys. Rev. D},
  volume = {35},
  issue = {12},
  pages = {3632--3636},
  numpages = {0},
  year = {1987},
  month = {Jun},
  publisher = {American Physical Society},
  doi = {10.1103/PhysRevD.35.3632},
  url = {https://link.aps.org/doi/10.1103/PhysRevD.35.3632}
}
@article{PhysRevD.68.124017,
  title = {Gravitational quasinormal radiation of higher-dimensional black holes},
  author = {Konoplya, R. A.},
  journal = {Phys. Rev. D},
  volume = {68},
  issue = {12},
  pages = {124017},
  numpages = {7},
  year = {2003},
  month = {Dec},
  publisher = {American Physical Society},
  doi = {10.1103/PhysRevD.68.124017},
  url = {https://link.aps.org/doi/10.1103/PhysRevD.68.124017}
}
@article{Konoplya:2004ip,
      author         = "Konoplya, R. A.",
      title          = "{Quasinormal modes of the Schwarzschild black hole and
                        higher order WKB approach}",
      journal        = "J. Phys. Stud.",
      volume         = "8",
      year           = "2004",
      pages          = "93-100",
      SLACcitation   = "
}
@article{PhysRevD.68.024018,
  title = {Quasinormal behavior of the $D$-dimensional Schwarzschild black hole and the higher order WKB approach},
  author = {Konoplya, R. A.},
  journal = {Phys. Rev. D},
  volume = {68},
  issue = {2},
  pages = {024018},
  numpages = {8},
  year = {2003},
  month = {Jul},
  publisher = {American Physical Society},
  doi = {10.1103/PhysRevD.68.024018},
  url = {https://link.aps.org/doi/10.1103/PhysRevD.68.024018}
}
@article{Ciftci_2003,
    doi = {10.1088/0305-4470/36/47/008},
    url = {https://doi.org/10.1088
    year = 2003,
    month = {nov},
    publisher = {{IOP} Publishing},
    volume = {36},
    number = {47},
    pages = {11807--11816},
    author = {Hakan Ciftci and Richard L Hall and Nasser Saad},
    title = {Asymptotic iteration method for eigenvalue problems},
    journal = {Journal of Physics A: Mathematical and General},
    abstract = {An asymptotic iteration method for solving second-order homogeneous linear differential equations of the form y″ = λ0(x)y′ + s0(x)y is introduced, where λ0(x) ≠ 0 and s0(x) are C∞ functions. Applications to Schrödinger-type problems, including some with highly singular potentials, are presented.}
}

@article{article, author = {Ciftci, Hakan and Hall, Richard and
Saad, Nasser},
year = {2005},
month = {06},
pages = {388-396},
title = {Perturbation theory in a framework of iteration methods},
volume = {340},
journal = {Phys Lett A},
doi =
{10.1016/j.physleta.2005.04.030} }

@article{Cho_2010,
    doi = {10.1088/0264-9381/27/15/155004},
    url = {https://doi.org/10.1088
    year = 2010,
    month = {jun},
    publisher = {{IOP} Publishing},
    volume = {27},
    number = {15},
    pages = {155004},
    author = {H T Cho and A S Cornell and Jason Doukas and Wade Naylor},
    title = {Black hole quasinormal modes using the asymptotic iteration method},
    journal = {Class. Quant. Grav.}
}
@Article{Bardeen,
author = { Bardeen,Jonh},
journal = {published in the conference proceeding in the U.S.S.R},
year = {1968},
OPTnote = {presented at GR5},
}

@Article{Ayon-Beato1999,
author="Ayon-Beato, Eloy and Garcia,
Alberto",
title="Non-Singular Charged Black Hole Solution for
Non-Linear Source",
journal="Gen. Relativ.  Gravit.",
year="1999",
month="May", day="01", volume="31",
number="5", pages="629--633",
abstract="A non-singular exact black hole solution inGeneral
Relativity is presented. The source is anon-linear electrodynamic
field, which reduces to theMaxwell theory for weak field. The
solution corresponds to a charged black hole with |q| ≤2scm ≈
0.6 m, having metric, curvatureinvariants, and electric field
boundedeverywhere.", issn="1572-9532",
doi="10.1023/A:1026640911319",
url="https://doi.org/10.1023/A:1026640911319" }

@article{AyonBeato:1998ub,
      author         = "Ayon-Beato, Eloy and Garcia, Alberto",
      title          = "{Regular black hole in general relativity coupled to
                        nonlinear electrodynamics}",
      journal        = "Phys. Rev. Lett.",
      volume         = "80",
      year           = "1998",
      pages          = "5056-5059",
      doi            = "10.1103/PhysRevLett.80.5056",
      eprint         = "gr-qc/9911046",
      archivePrefix  = "arXiv",
      primaryClass   = "gr-qc",
      SLACcitation   = "
}
@article{AyonBeato:1999rg,
      author         = "Ayon-Beato, Eloy and Garcia, Alberto",
      title          = "{New regular black hole solution from nonlinear
                        electrodynamics}",
      journal        = "Phys. Lett.",
      volume         = "B464",
      year           = "1999",
      pages          = "25",
      doi            = "10.1016/S0370-2693(99)01038-2",
      eprint         = "hep-th/9911174",
      archivePrefix  = "arXiv",
      primaryClass   = "hep-th",
      SLACcitation   = "
}
@article{Borde,
  title = {Open and closed universes, initial singularities, and inflation},
  author = {Borde, Arvind},
  journal = {Phys. Rev. D},
  volume = {50},
  issue = {6},
  pages = {3692--3702},
  numpages = {0},
  year = {1994},
  month = {Sep},
  publisher = {American Physical Society},
  doi = {10.1103/PhysRevD.50.3692},
  url = {https://link.aps.org/doi/10.1103/PhysRevD.50.3692}
}
@article{Barrab,
  title = {How many new worlds are inside a black hole?},
  author = {Barrab\`es, Claude and Frolov, Valeri P.},
  journal = {Phys. Rev. D},
  volume = {53},
  issue = {6},
  pages = {3215--3223},
  numpages = {0},
  year = {1996},
  month = {Mar},
  publisher = {American Physical Society},
  doi = {10.1103/PhysRevD.53.3215},
  url = {https://link.aps.org/doi/10.1103/PhysRevD.53.3215}
}
@article{Mars_1996,
    doi = {10.1088/0264-9381/13/5/003},
    url = {https://doi.org/10.1088
    year = 1996,
    month = {may},
    publisher = {{IOP} Publishing},
    volume = {13},
    number = {5},
    pages = {L51--L58},
    author = {Marc Mars and M Merc{\`{e}} Mart{\'{\i}}n-Prats and Jos{\'{e}} M M Senovilla},
    title = {Models of regular Schwarzschild black holes satisfying weak energy conditions},
    journal = {Class. Quant. Grav.},
    abstract = {We prove the existence of regular Schwarzschild black holes satisfying the weak energy conditions everywhere by presenting two explicit models. One of these models is explicitly seen to be complete (and therefore regular) by giving a maximal extension across the horizons.}
}

 @Article{Maldacena1999,
author="Maldacena, Juan", title="The Large-N Limit of
Superconformal Field Theories and Supergravity", journal="Int. J.
Theor. Phys.",
year="1999",
month="Apr",
day="01",
volume="38",
number="4",
pages="1113--1133",
abstract="We show that the large-N limits of certainconformal field theories in various dimensions includein their Hilbert space a sector describing supergravityon the product of anti-de Sitter spacetimes, spheres, and other compact manifolds. This is shown bytaking some branes in the full M/string theory and thentaking a low-energy limit where the field theory on thebrane decouples from the bulk. We observe that, in this limit, we can still trust thenear-horizon geometry for large N. The enhancedsupersymmetries of the near-horizon geometry correspondto the extra supersymmetry generators present in thesuperconformal group (as opposed to just the super-Poincaregroup). The 't Hooft limit of 3 + 1 N = 4 super-Yang--Mills at the conformal pointis shown to contain strings: they are IIB strings. Weconjecture that compactifications of M/string theory on various anti-de Sitterspacetimes is dual to various conformal field theories.This leads to a new proposal for a definition ofM-theory which could be extended to include fivenoncompact dimensions.",
issn="1572-9575",
doi="10.1023/A:1026654312961",
url="https://doi.org/10.1023/A:1026654312961"
}


@article{PhysRevD.62.024027,
  title = {Quasinormal modes of AdS black holes and the approach to thermal equilibrium},
  author = {Horowitz, Gary T. and Hubeny, Veronika E.},
  journal = {Phys. Rev. D},
  volume = {62},
  issue = {2},
  pages = {024027},
  numpages = {11},
  year = {2000},
  month = {Jun},
  publisher = {American Physical Society},
  doi = {10.1103/PhysRevD.62.024027},
  url = {https://link.aps.org/doi/10.1103/PhysRevD.62.024027}
}
@article{BEKENSTEIN19957,
title = "Spectroscopy of the quantum
black hole", journal = "Phys. Lett. B",
volume = "360",
number = "1",
pages = "7 - 12",
year = "1995",
issn = "0370-2693",
doi = "https://doi.org/10.1016/0370-2693(95)01148-J",
url = "http://www.sciencedirect.com/science/article/pii/037026939501148J",
author = "Jacob D. Bekenstein and V.F. Mukhanov",
abstract = "We develop the idea that, in quantum gravity where the horizon fluctuates, a black hole should have a discrete mass spectrum with concomitant line emission. Simple arguments fix the spacing of the lines, which should be broad but unblended. Assuming uniformity of the matrix elements for quantum transitions between near levels, we work out the probabilities for the emission of a specified series of quanta and the intensities of the spectral lines. The thermal character of the radiation is entirely due to the degeneracy of the levels, the same degeneracy that becomes manifest as black hole entropy. One prediction is that there should be no lines with wavelength of the order of the black hole size or larger. This makes it possible to test quantum gravity with black holes well above Planck scale."
}


@article{PhysRevLett.81.4293,
  title = {Bohr's Correspondence Principle and the Area Spectrum of Quantum Black Holes},
  author = {Hod, Shahar},
  journal = {Phys. Rev. Lett.},
  volume = {81},
  issue = {20},
  pages = {4293--4296},
  numpages = {0},
  year = {1998},
  month = {Nov},
  publisher = {American Physical Society},
  doi = {10.1103/PhysRevLett.81.4293},
  url = {https://link.aps.org/doi/10.1103/PhysRevLett.81.4293}
}
@article{PhysRevLett.90.081301,
  title = {Quasinormal Modes, the Area Spectrum, and Black Hole Entropy},
  author = {Dreyer, Olaf},
  journal = {Phys. Rev. Lett.},
  volume = {90},
  issue = {8},
  pages = {081301},
  numpages = {4},
  year = {2003},
  month = {Feb},
  publisher = {American Physical Society},
  doi = {10.1103/PhysRevLett.90.081301},
  url = {https://link.aps.org/doi/10.1103/PhysRevLett.90.081301}
}
@article{PhysRevD.2.2141,
  title = {Gravitational Field of a Particle Falling in a Schwarzschild Geometry Analyzed in Tensor Harmonics},
  author = {Zerilli, Frank J.},
  journal = {Phys. Rev. D},
  volume = {2},
  issue = {10},
  pages = {2141--2160},
  numpages = {0},
  year = {1970},
  month = {Nov},
  publisher = {American Physical Society},
  doi = {10.1103/PhysRevD.2.2141},
  url = {https://link.aps.org/doi/10.1103/PhysRevD.2.2141}
}

@article{PhysRevLett.24.737,
  title = {Effective Potential for Even-Parity Regge-Wheeler Gravitational Perturbation Equations},
  author = {Zerilli, Frank J.},
  journal = {Phys. Rev. Lett.},
  volume = {24},
  issue = {13},
  pages = {737--738},
  numpages = {0},
  year = {1970},
  month = {Mar},
  publisher = {American Physical Society},
  doi = {10.1103/PhysRevLett.24.737},
  url = {https://link.aps.org/doi/10.1103/PhysRevLett.24.737}
}
@article{PhysRevD.1.2870,
  title = {Stability of the Schwarzschild Metric},
  author = {Vishveshwara, C. V.},
  journal = {Phys. Rev. D},
  volume = {1},
  issue = {10},
  pages = {2870--2879},
  numpages = {0},
  year = {1970},
  month = {May},
  publisher = {American Physical Society},
  doi = {10.1103/PhysRevD.1.2870},
  url = {https://link.aps.org/doi/10.1103/PhysRevD.1.2870}
}
@article{Vishveshwara:1970zz,
      author         = "Vishveshwara, C. V.",
      title          = "{Scattering of Gravitational Radiation by a Schwarzschild
                        Black-hole}",
      journal        = "Nature",
      volume         = "227",
      year           = "1970",
      pages          = "936-938",
      doi            = "10.1038/227936a0",
      SLACcitation   = "
}
@article{Chandrasekhar:1975zza,
      author         = "Chandrasekhar, S. and Detweiler, Steven L.",
      title          = "{The quasi-normal modes of the Schwarzschild black hole}",
      journal        = "Proc. Roy. Soc. Lond.",
      volume         = "A344",
      year           = "1975",
      pages          = "441-452",
      doi            = "10.1098/rspa.1975.0112",
      SLACcitation   = "
}
@article{BLOME1984231,
title = "Quasi-normal oscillations of a schwarzschild black hole",
journal = "Physics Letters A",
volume = "100",
number = "5",
pages = "231 - 234",
year = "1984",
issn = "0375-9601",
doi = "https://doi.org/10.1016/0375-9601(84)90769-2",
url = "http://www.sciencedirect.com/science/article/pii/0375960184907692",
author = "Hans-Joachim Blome and Bahram Mashhoon",
abstract = "The modes of oscillation of a Schwarzschild black hole are determined within an analytic framework. These quasi-normal modes are related to the bound states of the inverted black hole potential which is approximated by the inverted Eckart potential. For a given angular momentum parameter j, the real part of the quasi-normal frequency decreases as the mode number n (or, equivalently, the damping factor) increases, in agreement with the results of numerical studies."
}
@article{PhysRevD.30.295,
  title = {New approach to the quasinormal modes of a black hole},
  author = {Ferrari, Valeria and Mashhoon, Bahram},
  journal = {Phys. Rev. D},
  volume = {30},
  issue = {2},
  pages = {295--304},
  numpages = {0},
  year = {1984},
  month = {Jul},
  publisher = {American Physical Society},
  doi = {10.1103/PhysRevD.30.295},
  url = {https://link.aps.org/doi/10.1103/PhysRevD.30.295}
}

@article{doi:10.1063/1.527130, author = {Leaver,E. W. }, title =
{Solutions to a generalized spheroidal wave equation:
Teukolsky’s equations in general relativity, and the
two‐center problem in molecular quantum mechanics}, journal =
{J. Math. Phys.}, volume = {27}, number = {5}, pages =
{1238-1265}, year = {1986}, doi = {10.1063/1.527130},

URL = {
        https://doi.org/10.1063/1.527130

},
eprint = {
        https://doi.org/10.1063/1.527130

}}
@article{PhysRevD.34.384,
  title = {Spectral decomposition of the perturbation response of the Schwarzschild geometry},
  author = {Leaver, Edward W.},
  journal = {Phys. Rev. D},
  volume = {34},
  issue = {2},
  pages = {384--408},
  numpages = {0},
  year = {1986},
  month = {Jul},
  publisher = {American Physical Society},
  doi = {10.1103/PhysRevD.34.384},
  url = {https://link.aps.org/doi/10.1103/PhysRevD.34.384}
}
@article{Leaver:1985ax,
      author         = "Leaver, E. W.",
      title          = "{An Analytic representation for the quasi normal modes of
                        Kerr black holes}",
      journal        = "Proc. Roy. Soc. Lond.",
      volume         = "A402",
      year           = "1985",
      pages          = "285-298",
      doi            = "10.1098/rspa.1985.0119"
}
@article{Berti_2009,
    doi = {10.1088/0264-9381/26/16/163001},
    url = {https://doi.org/10.1088
    year = 2009,
    month = {jul},
    publisher = {{IOP} Publishing},
    volume = {26},
    number = {16},
    pages = {163001},
    author = {Emanuele Berti and Vitor Cardoso and Andrei O Starinets},
    title = {Quasinormal modes of black holes and black branes},
    journal = {Class. Quant. Grav.},
    abstract = {Quasinormal modes are eigenmodes of dissipative systems. Perturbations of classical gravitational backgrounds involving black holes or branes naturally lead to quasinormal modes. The analysis and classification of the quasinormal spectra require solving non-Hermitian eigenvalue problems for the associated linear differential equations. Within the recently developed gauge-gravity duality, these modes serve as an important tool for determining the near-equilibrium properties of strongly coupled quantum field theories, in particular their transport coefficients, such as viscosity, conductivity and diffusion constants. In astrophysics, the detection of quasinormal modes in gravitational wave experiments would allow precise measurements of the mass and spin of black holes as well as new tests of general relativity. This review is meant as an introduction to the subject, with a focus on the recent developments in the field.}
}
@Article{Kokkotas1999,
author="Kokkotas, Kostas D.
and Schmidt, Bernd G.",
title="Quasi-Normal Modes of Stars and Black Holes",
journal="Living Reviews in Relativity",
year="1999",
month="Sep",
day="16",
volume="2",
number="1",
pages="2",
abstract="Perturbations of stars and black holes have been one of the main topics of relativistic astrophysics for the last few decades. They are of particular importance today, because of their relevance to gravitational wave astronomy. In this review we present the theory of quasi-normal modes of compact objects from both the mathematical and astrophysical points of view. The discussion includes perturbations of black holes (Schwarzschild, Reissner-Nordstr{\"o}m, Kerr and Kerr-Newman) and relativistic stars (non-rotating and slowly-rotating). The properties of the various families of quasi-normal modes are described, and numerical techniques for calculating quasi-normal modes reviewed. The successes, as well as the limits, of perturbation theory are presented, and its role in the emerging era of numerical relativity and supercomputers is discussed.",
issn="1433-8351",
doi="10.12942/lrr-1999-2",
url="https://doi.org/10.12942/lrr-1999-2"
}
@article{Nollert:1999ji,
      author         = "Nollert, Hans-Peter",
      title          = "{TOPICAL REVIEW: Quasinormal modes: the characteristic
                        `sound' of black holes and neutron stars}",
      journal        = "Class. Quant. Grav.",
      volume         = "16",
      year           = "1999",
      pages          = "R159-R216",
      doi            = "10.1088/0264-9381/16/12/201",
      SLACcitation   = "
}

@Article{Mashh,
author= {Barham Mashhoon},
journal = {Proc. 3rd Marcel Grossmann Meeting on Recent Developments of General Relativity
ed H Ning (Amsterdam: North-Holland)},
year = {1983},
page={p 599}
}

@ARTICLE{1972ApJ...178..347B,
   author = {{Bardeen}, J.~M. and {Press}, W.~H. and {Teukolsky}, S.~A.},
    title = "{Rotating Black Holes: Locally Nonrotating Frames, Energy Extraction, and Scalar Synchrotron Radiation}",
  journal = {Astrophys. J},
     year = 1972,
    month = dec,
   volume = 178,
    pages = {347-370},
      doi = {10.1086/151796},
   adsurl = {https://ui.adsabs.harvard.edu/abs/1972ApJ...178..347B},
  adsnote = {Provided by the SAO/NASA Astrophysics Data System}
}
@article{PhysRevD.87.024034,
  title = {Quasinormal modes of regular black holes},
  author = {Flachi, Antonino and Lemos, Jos\'e P. S.},
  journal = {Phys. Rev. D},
  volume = {87},
  issue = {2},
  pages = {024034},
  numpages = {8},
  year = {2013},
  month = {Jan},
  publisher = {American Physical Society},
  doi = {10.1103/PhysRevD.87.024034},
  url = {https://link.aps.org/doi/10.1103/PhysRevD.87.024034}
}
@article{PhysRevD.91.083008,
  title = {Quasinormal modes of test fields around regular black holes},
  author = {Toshmatov, Bobir and Abdujabbarov, Ahmadjon and Stuchl\'{\i}k, Zden\ifmmode \check{e}\else \v{e}\fi{}k and Ahmedov, Bobomurat},
  journal = {Phys. Rev. D},
  volume = {91},
  issue = {8},
  pages = {083008},
  numpages = {13},
  year = {2015},
  month = {Apr},
  publisher = {American Physical Society},
  doi = {10.1103/PhysRevD.91.083008},
  url = {https://link.aps.org/doi/10.1103/PhysRevD.91.083008}
}
@article{Li:2016oif,
      author         = "Li, Jin and Lin, Kai and Wen, Hao",
      title          = "{Gravitational Quasinormal Modes of Regular Phantom Black
                        Hole}",
      journal        = "Adv. High Energy Phys.",
      volume         = "2017",
      year           = "2017",
      pages          = "5234214",
      doi            = "10.1155/2017/5234214",
      eprint         = "1605.08502",
      archivePrefix  = "arXiv",
      primaryClass   = "gr-qc",
      SLACcitation   = "
}


@article{Lopez:2018aec,
    author = "Lopez, L. A. and Hinojosa, Valeria",
    title = "{Quasinormal modes of Charged Regular Black Hole}",
    eprint = "1810.09034",
    archivePrefix = "arXiv",
    primaryClass = "gr-qc",
    doi = "10.1139/cjp-2019-0572",
    journal = "Can. J. Phys.",
    volume = "99",
    number = "1",
    pages = "44--48",
    year = "2021"
}
@article{Toshmatov:2018ell,
      author         = "Toshmatov, Bobir and Stuchlík, Zdeněk and Ahmedov,
                        Bobomurat",
      title          = "{Electromagnetic perturbations of black holes in general
                        relativity coupled to nonlinear electrodynamics: Polar
                        perturbations}",
      journal        = "Phys. Rev.",
      volume         = "D98",
      year           = "2018",
      number         = "8",
      pages          = "085021",
      doi            = "10.1103/PhysRevD.98.085021",
      eprint         = "1810.06383",
      archivePrefix  = "arXiv",
      primaryClass   = "gr-qc",
      SLACcitation   = "
}

@article{ refId0,
	author = {{Wu, Chen}},
	title = {Quasinormal frequencies of gravitational perturbation in regular black hole spacetimes},
	DOI= "10.1140/epjc/s10052-018-5764-6",
	url= "https://doi.org/10.1140/epjc/s10052-018-5764-6",
	journal = {Eur. Phys. J. C},
	year = 2018,
	volume = 78,
	number = 4,
	pages = "283",
}

@article{PhysRevD.63.044005,
  title = {Regular magnetic black holes and monopoles from nonlinear electrodynamics},
  author = {Bronnikov, K. A.},
  journal = {Phys. Rev. D},
  volume = {63},
  issue = {4},
  pages = {044005},
  numpages = {6},
  year = {2001},
  month = {Jan},
  publisher = {American Physical Society},
  doi = {10.1103/PhysRevD.63.044005},
  url = {https://link.aps.org/doi/10.1103/PhysRevD.63.044005}
}


@article{Berej:2006cc,
    author = "Berej, Waldemar and Matyjasek, Jerzy and Tryniecki, Dariusz and Woronowicz, Mariusz",
    title = "{Regular black holes in quadratic gravity}",
    eprint = "hep-th/0606185",
    archivePrefix = "arXiv",
    doi = "10.1007/s10714-006-0270-9",
    journal = "Gen. Rel. Grav.",
    volume = "38",
    pages = "885--906",
    year = "2006"
}

@article{Ayon-Beato:2000mjt,
    author = "Ayon-Beato, Eloy and Garcia, Alberto",
    title = "{The Bardeen model as a nonlinear magnetic monopole}",
    eprint = "gr-qc/0009077",
    archivePrefix = "arXiv",
    doi = "10.1016/S0370-2693(00)01125-4",
    journal = "Phys. Lett. B",
    volume = "493",
    pages = "149--152",
    year = "2000"
}

@article{PhysRevLett.96.031103,
  title = {Formation and Evaporation of Nonsingular Black Holes},
  author = {Hayward, Sean A.},
  journal = {Phys. Rev. Lett.},
  volume = {96},
  issue = {3},
  pages = {031103},
  numpages = {4},
  year = {2006},
  month = {Jan},
  publisher = {American Physical Society},
  doi = {10.1103/PhysRevLett.96.031103},
  url = {https://link.aps.org/doi/10.1103/PhysRevLett.96.031103}
}
@article{Panotopoulos:2019qjk,
    author = "Panotopoulos, Grigoris and Rinc\'on, \'Angel",
    title = "{Quasinormal modes of regular black holes with non linear-Electrodynamical sources}",
    eprint = "1904.10847",
    archivePrefix = "arXiv",
    primaryClass = "gr-qc",
    doi = "10.1140/epjp/i2019-12686-x",
    journal = "Eur. Phys. J. Plus",
    volume = "134",
    number = "6",
    pages = "300",
    year = "2019"
}
@article{Churilova:2019cyt,
    author = "Churilova, M. S. and Stuchlik, Z.",
    title = "{Ringing of the regular black-hole/wormhole transition}",
    eprint = "1911.11823",
    archivePrefix = "arXiv",
    primaryClass = "gr-qc",
    doi = "10.1088/1361-6382/ab7717",
    journal = "Class. Quant. Grav.",
    volume = "37",
    number = "7",
    pages = "075014",
    year = "2020"
}

@article{PhysRevD.101.064004,
  title = {Echoes in brane worlds: Ringing at a black hole-wormhole transition},
  author = {Bronnikov, Kirill A. and Konoplya, Roman A.},
  journal = {Phys. Rev. D},
  volume = {101},
  issue = {6},
  pages = {064004},
  numpages = {9},
  year = {2020},
  month = {Mar},
  publisher = {American Physical Society},
  doi = {10.1103/PhysRevD.101.064004},
  url = {https://link.aps.org/doi/10.1103/PhysRevD.101.064004}
}
@article{PhysRevD.103.064016,
  title = {Physical properties of a regular rotating black hole: Thermodynamics, stability, and quasinormal modes},
  author = {Hendi, S. H. and Sajadi, S. N. and Khademi, M.},
  journal = {Phys. Rev. D},
  volume = {103},
  issue = {6},
  pages = {064016},
  numpages = {16},
  year = {2021},
  month = {Mar},
  publisher = {American Physical Society},
  doi = {10.1103/PhysRevD.103.064016},
  url = {https://link.aps.org/doi/10.1103/PhysRevD.103.064016}
}
@article{LAN2021115539,
title = {Quasinormal modes and phase transitions of regular black holes},
journal = {Nuclear Physics B},
volume = {971},
pages = {115539},
year = {2021},
issn = {0550-3213},
doi = {https://doi.org/10.1016/j.nuclphysb.2021.115539},
url = {https://www.sciencedirect.com/science/article/pii/S0550321321002364},
author = {Chen Lan and Yan-Gang Miao and Hao Yang},
abstract = {By applying the dimensionless scheme, we investigate the quasinormal modes and phase transitions analytically for three types of regular black holes. The universal deviations to the first law of mechanics in regular black holes are proved. Meanwhile, we verify that second order phase transitions and Davies points still exist in these three models. In addition, we calculate their quasinormal modes in the eikonal limit by applying the light ring/quasinormal mode correspondence, and discuss the spiral-like shapes and the relations between the quasinormal modes and phase transitions. As the main result, we show that spiral-like shapes in the complex frequency plane are closely related to the parameterization, namely in some particular units the spiral-like shapes will emerge in the models, which may not be of the spiral behaviors reported by other authors. We also discover a universal property of regular black holes, i.e., the imaginary parts of their QNMs do not vanish for the extreme cases, which does not appear in singular black holes, such as the Reissner-Nordström and Kerr black holes, etc.}
}
@article{Cai:2020kue,
    author = "Cai, Xin-Chang and Miao, Yan-Gang",
    title = "{Quasinormal modes of the generalized Ay\'on-Beato\textendash{}Garc\'\i{}a black hole in scalar-tensor-vector gravity}",
    eprint = "2008.04576",
    archivePrefix = "arXiv",
    primaryClass = "gr-qc",
    doi = "10.1103/PhysRevD.102.084061",
    journal = "Phys. Rev. D",
    volume = "102",
    number = "8",
    pages = "084061",
    year = "2020"
}
@article{Bronnikov:2012ch,
    author = "Bronnikov, K. A. and Konoplya, R. A. and Zhidenko, A.",
    title = "{Instabilities of wormholes and regular black holes supported by a phantom scalar field}",
    eprint = "1205.2224",
    archivePrefix = "arXiv",
    primaryClass = "gr-qc",
    doi = "10.1103/PhysRevD.86.024028",
    journal = "Phys. Rev. D",
    volume = "86",
    pages = "024028",
    year = "2012"
}
@article{PhysRevD.103.124050,
  title = {Quasinormal modes and shadows of a new family of Ay\'on-Beato-Garc\'{\i}a black holes},
  author = {Cai, Xin-Chang and Miao, Yan-Gang},
  journal = {Phys. Rev. D},
  volume = {103},
  issue = {12},
  pages = {124050},
  numpages = {16},
  year = {2021},
  month = {Jun},
  publisher = {American Physical Society},
  doi = {10.1103/PhysRevD.103.124050},
  url = {https://link.aps.org/doi/10.1103/PhysRevD.103.124050}
}


\begin{thebibliography}{}
\bibitem{Maldacena1999}J Maldacena \textit{Int.J.Theor.Phys.} {\bf 38}, 1113(1999)
\bibitem{PhysRevD.62.024027}G T Horowitz and V E Hubeny \textit{Phys. Rev. D} {\bf 62}, 024027(2000)
\bibitem{BEKENSTEIN19957}J D Bekenstein and V Mukhanov \textit{Phys. Lett. B} {\bf 360}, 7(1995)
\bibitem{PhysRevLett.81.4293}S Hod \textit{Phys. Rev. Lett.} {\bf 81}, 4293(1998)
\bibitem{PhysRevLett.90.081301}O Dreyer \textit{Phys. Rev. Lett.} {\bf 90}, 081301(2003)
\bibitem{Regge:1957td}T Regge and J A Wheeler \textit{Phys. Rev.} {\bf 108}, 1063(1957)
\bibitem{PhysRevD.2.2141}F J Zerilli \textit{Phys. Rev. D} {\bf 2}, 2141(1970)
\bibitem{PhysRevLett.24.737}F J Zerilli \textit{Phys. Rev. Lett.} {\bf 24}, 737(1970)
\bibitem{PhysRevD.1.2870}C V Vishveshwara \textit{Phys. Rev. D} {\bf 1}, 2870(1970)
\bibitem{Vishveshwara:1970zz}C V Vishveshwara \textit{Nature} {\bf 227}, 936(1970)
\bibitem{Chandrasekhar:1975zza}S Chandrasekhar and S L Detweiler \textit{Proc. Roy. Soc. Lond.} {\bf A344}, 441(1975)
\bibitem{BLOME1984231}H J Blome and B Mashhoon \textit{Phys. Lett. A} {\bf 100}, 231(1984)
\bibitem{PhysRevD.30.295}V Ferrari and B Mashhoon \textit{Phys. Rev. D} {\bf 30}, 295(1984)
\bibitem{doi:10.1063/1.527130}E W Leaver \textit{J. Math. Phys.} {\bf 27}, 1238(1986)
\bibitem{PhysRevD.34.384}E W Leaver \textit{Phys. Rev. D} {\bf 34}, 384(1986)
\bibitem{Leaver:1985ax}E W Leaver \textit{Proc. Roy. Soc. Lond.} {\bf A402}, 285(1985)
\bibitem{Horowitz-Hubeny:2000}G T Horowitz and V E Hubeny \textit{Phys. Rev. D} {\bf 62}, 024027(2000)
\bibitem{Cho_2010}H T Cho, A S Cornell, J Doukas, and W Naylor \textit{Class. Quant. Grav.} {\bf 27}, 155004(2010)
\bibitem{Nollert:1999ji}H P Nollert \textit{Class. Quant. Grav.} {\bf 16}, 159(1999)
\bibitem{Kokkotas1999}K D Kokkotas and B G Schmidt \textit{Living Rev. Relativ} {\bf 2}, 2(1999)
\bibitem{Berti_2009}E Berti, V Cardoso, and A O Starinets \textit{Class. Quant. Grav.} {\bf 26}, 163001(2009)
\bibitem{Konoplya:2011qq}R A Konoplya and A Zhidenko \textit{Rev. Mod. Phys.} {\bf 83}, 793(2011)
\bibitem{Bardeen}J Bardeen \textit{published in the conference proceeding in the U.S.S.R} (1968)
\bibitem{Mars_1996}M Mars, M M Mart{\'{\i}}n-Prats, and J M M Senovilla \textit{Class. Quant. Grav.} {\bf 13}, 51(1996)
\bibitem{Barrab}C Barrab\`es and V P Frolov \textit{Phys. Rev. D} {\bf 53}, 3215(1996)
\bibitem{Borde}A Borde \textit{Phys. Rev. D} {\bf 50}, 3692(1994)
\bibitem{AyonBeato:1998ub}E Ayon-Beato and A Garcia \textit{Phys. Rev. Lett.} {\bf 80}, 5056(1998)
\bibitem{Ayon-Beato1999}E Ayon-Beato and A Garcia \textit{Gen. Relativ. Gravit.} {\bf 31}, 629(1999)
\bibitem{Ayon-Beato:2000mjt}E Ayon-Beato and A Garcia \textit{Phys. Lett. B} {\bf 493}, 149(2000)
\bibitem{PhysRevLett.96.031103}S A Hayward \textit{Phys. Rev. Lett.} {\bf 96}, 031103(2006)
\bibitem{PhysRevD.87.024034}A Flachi and J P S Lemos \textit{Phys. Rev. D} {\bf 87}, 024034(2013)
\bibitem{PhysRevD.91.083008}B Toshmatov, A Abdujabbarov, Z  Stuchl\'{\i}k, and B Ahmedov \textit{Phys. Rev. D} {\bf 91}, 083008(2015)
\bibitem{Toshmatov:2018ell}B Toshmatov, Z Stuchl\'{\i}k, and B Ahmedov \textit{Phys. Rev. D} {\bf 98}, 085021(2018)
\bibitem{refId0}C Wu \textit{Eur. Phys. J. C} {\bf 78}, 283(2018)
\bibitem{Li:2016oif}J Li, K Lin, and H Wen \textit{Adv. High Energy Phys.} {\bf 2017}, 5234214(2017)
\bibitem{Lopez:2018aec}L A Lopez and V Hinojosa \textit{Can. J. Phys.} {\bf 99(1)}, 44(2020)
\bibitem{Panotopoulos:2019qjk}G Panotopoulos and A Rinc\'on \textit{Eur. Phys. J. Plus} {\bf 134}, 300(2019)
\bibitem{PhysRevD.103.124050}X C Cai and Y G Miao \textit{Phys. Rev. D} {\bf 103}, 124050(2021)
\bibitem{PhysRevD.103.064016}S H Hendi, S N Sajadi, and M Khademi \textit{Phys. Rev. D} {\bf 103}, 064016(2021)
\bibitem{Cai:2020kue}X C Cai and Y G Miao \textit{Phys. Rev. D} {\bf 102}, 084061(2020)
\bibitem{Bronnikov:2012ch}K A Bronnikov, R A Konoplya and A Zhidenko \textit{Phys. Rev. D} {\bf 86}, 024028(2012)
\bibitem{Churilova:2019cyt}M S Churilova and Z Stuchlik \textit{Class. Quant. Grav.} {\bf 37}, 075014(2020)
\bibitem{PhysRevD.101.064004}K A Bronnikov and R A Konoplya \textit{Phys. Rev. D} {\bf 101}, 064004(2020)
\bibitem{AyonBeato:1999rg}E Ayon-Beato and A Garcia \textit{Phys. Lett. B} {\bf 464}, 25(1999)
\bibitem{Berej:2006cc}W Berej, J Matyjasek, D Tryniecki and M Woronowicz \textit{Gen. Rel. Grav.} {\bf 38}, 885(2006)
\bibitem{PhysRevD.63.044005}K A Bronnikov \textit{Phys. Rev. D} {\bf 63}, 044005(2001)
\bibitem{1972ApJ...178..347B}J M Bardeen, W H Press, and S A Teukolsky \textit{Astrophys. J} {\bf 178}, 347(1972)
\bibitem{Mashh}B. Mashhoon \textit{Proc. 3rd Marcel Grossmann Meeting on Recent Developments of General Relativity ed H Ning (Amsterdam: North-Holland)}, (1983)
\bibitem{Schutz:1985}B F Schutz and C M Will \textit{Astrophys. J} {\bf 291}, 33(1985)
\bibitem{Iyer-I}S Iyer and C M Will \textit{Phys. Rev. D} {\bf 35}, 3621(1987)
\bibitem{PhysRevD.68.124017}R A Konoplya \textit{Phys. Rev. D} {\bf 68}, 124017(2003)
\bibitem{Konoplya:2004ip}R A Konoplya \textit{J. Phys. Stud.} {\bf 8}, 93(2004)
\bibitem{Iyer-II}S Iyer \textit{Phys. Rev. D} {\bf 35}, 3632(1987)
\bibitem{PhysRevD.68.024018}R A Konoplya \textit{Phys. Rev. D} {\bf 68}, 024018(2003)
\bibitem{Ciftci_2003}H Ciftci, R L Hall and N Saad \textit{J. Phys. Math. Gen.} {\bf 36}, 11807(2003)
\bibitem{article}H Ciftci, R Hall and N Saad \textit{Phys. Lett. A} {\bf 340}, 388(2005)
\end{thebibliography}
\end{document}